\documentclass[reprint, amsmath,amssymb,aps, onecolumn]{revtex4-2}

\usepackage{graphicx}
\usepackage{subfigure}
\usepackage{subfloat}
\usepackage{amssymb}
\usepackage{amsmath}
\usepackage{multirow}
\usepackage{dcolumn}
\usepackage{bm}
\usepackage{wasysym}
\usepackage{cancel}

\usepackage{txfonts}
\usepackage{lineno}

\begin{document}

\title{Analytical approach to the design of RF photoinjector}

\author{Kiwan Park}
 \affiliation{Department of Physics, Soongsil University, 369, Sangdo-ro, Dongjak-gu, Seoul 06978 Republic of Korea\\ pkiwan@ssu.ac.kr\\}
%
%

\date{\today}

\begin{abstract}
The objective of this thesis is to ascertain the dimensions of an RF 2.856GHz photoinjector through a combination of analytical and computational approaches. The phase velocity within a single cavity exceeds 'c', rendering it inadequate for storing the requisite energy for beam acceleration. To surmount this limitation, we aim to devise a multi-celled cavity design. However, the alterations in electromagnetic fields and resonant frequency within the multi-celled cavity are intricate and sensitive, presenting challenges in obtaining precise dimensions solely via computer simulations. Prior to numerical methods, it is essential to analyze the photoinjector using theoretical frameworks. We employ perturbation theory and the construction of an equivalent circuit to elucidate the underlying physics of the photoinjector and the electrical oscillations within the cell structure. Detailed analytical methods for the equivalent circuit are explored. Through theoretical analysis, the dimensions and simulation outcomes can be determined quantitatively.
\end{abstract}

\maketitle

\section{Introduction}
The availability of a high-brightness electron source is a critical necessity for practical applications such as Free-Electron Lasers (FEL), coherent electromagnetic wave generation, and wakefield accelerators(\cite{2022PhRvS..25f3401F}, and references therein). Various types of injectors, including traditional thermal DC, thermal RF, and the newer RF photoinjector, have been developed and utilized. However, the first two thermal injectors fail to meet the stringent requirements of many present and future applications due to their insufficient power. Consequently, the need arose for the development of a new type of photoinjector. In this introductory chapter, we provide brief explanations of the three types of injectors and their limitations\footnote{This work is based on the author's master thesis \citep{park2005}. All work was carried out at the Pohang Accelerator Laboratory (PAL).}.

\subsection{Evolution of Electron Beam Injectors}
The first type of injector employed a thermal cathode as an electron generator with a DC gun and RF buncher. A high DC voltage is applied to accelerate the electrons emitted from the thermal cathode. These electrons enter the RF buncher continuously, akin to an electron stream. This electron stream experiences an oscillating electric field. While some electrons accelerate during the positive half of the RF phase, others in the negative voltage phase experience a backward force, resulting in the formation of a beam bunch at the frequency of the RF field. However, despite its simple principle and structure, the thermal injector has a critical limitation. The force that accelerates the electrons from the cathode is relatively weak. Due to electron repulsion, the beam emittance expands rapidly (known as the space charge effect). To mitigate this emittance degradation, the beam must be accelerated as close to the speed of light as possible. Consequently, a high DC voltage between the anode and cathode is necessary. However, as the accelerating DC voltage increases, the likelihood of voltage breakdown between the cathode and anode also increases proportionally. Naturally, the beam current cannot be strengthened as much as desired. Therefore, this type of injector is not suitable for high beam bunch output.

\subsubsection{Thermal RF Injector}
In this setup, the DC positive anode is removed, and instead, an RF cavity incorporates the thermal cathode at its end wall. In this structure, the voltage gradient in the cavity approaches 100 MV/m. This injector can accelerate the beam bunch closer to the speed of light in a shorter time compared to the thermal DC injector. However, the thermal RF injector still exhibits some shortcomings. Firstly, due to the oscillation of the field, electrons emitted from the cathode during the negative phase may not advance forward. Even electrons that begin just before the negative phase may move backward and collide with the cathode. This phenomenon is referred to as "back-bombardment." While occasional back-bombardment may not pose a significant issue, repeated occurrences can lead to surface currents generated by electrons hitting the back wall, significantly affecting the field shape in the cavity. Moreover, continuous back-bombardment can cause the cathode to overheat and potentially become damaged. In summary, this structure fails to effectively separate the wave in the negative phase from the beam bunch. Secondly, the peak current is constrained by the thermal cathode current density, with the maximum density known to be 100 A/$cm^{2}$. These limitations prompted the development of a new injector: the RF photoinjector.

\subsubsection{RF Photoinjector with a single Cavity}
The photoinjector depicts the basic structure of a photoinjector, which appears similar to the thermal RF injector, except for the cathode attached to the back wall of the cavity. The cathode is illuminated by laser pulses, causing beam bunches to be emitted via the photoelectric effect. The principle of beam bunch motion is analogous to that of the thermal RF injector. This type of injector possesses three intrinsic properties. Firstly, it emits electrons according to the laser pulses. Since the laser pulse frequency is short (about a few picoseconds), the bunch length is so short that nearly all electrons in a bunch experience the same field phase. Secondly, there is no "back bombardment" as long as the laser illuminates at the appropriate RF phase frequency. The third advantage is the high current in bunches. The maximum current density is determined by the number of electrons emitted from a cathode. The current density of the photoinjector reaches as high as $21 kA/cm^2$, which is two hundred times greater than that of a thermal injector.

\subsubsection{RF Photoinjector Combined by Two Cavities}
This setup includes two cavities referred to as `a cell'. The rationale behind using multiple cavities is twofold. First, in a single cavity, the phase velocity of the wave exceeds that of light. Consequently, the beam bunch, whose velocity is below c, is overtaken by the field. This results in the beam bunch experiencing forces in both forward and backward directions, leading to a net zero accelerating force and no acceleration of the beam bunch. To address this, it is essential to reduce the phase velocity below c by introducing irises in the cavity, thereby dividing it into several cells. Second, there are other significant reasons for incorporating irises. They render the $E_{r}$ and $H_{\theta}$ fields linearly dependent on the distance $r$ from the axis, which can be controlled using an external solenoid. Additionally, they enhance energy storage without requiring additional facilities such as input couplers or higher voltage inputs. Waves enter the full cavity through an aperture. Since the wave's amplitude significantly decreases after passing through the aperture, the conductor wall separating the rectangular waveguide and the cavity must be very thin. However, the formation of an electric dipole at the thin edge of the aperture increases the risk of voltage breakdown. To mitigate this issue, attaching a rectangular waveguide appropriately is necessary. Based on conductivity, there are two types of RF cavities: superconducting RF (SCRF or SRF) and normal conducting RF (NCRF). SCRF is utilized for high-repetition-rate operations with low electric field at the cathode, resulting in increased transverse emittance\citep{2011PhRvS..14b4801A, 2021PhRvS..24c3401T}. Conversely, NCRF operates with high electric field rates, generating high-brightness electron beams\citep{2022PhRvS..25f3401F}.\\

We convert the 1.6-cell photoinjector into an equivalent circuit. In the following section, we provide a review of the basic background theory. Next, we construct the equivalent circuit and solve it using Kirchoff's loop rule. In the subsequent section, we present our simulation results. Finally, we conclude by summarizing our work.

\section{Background Knowledge: Understanding the Fundamentals}

\subsection{Basic circuit theory}
An injector typically consists of a conductor with specific lengths and surfaces interacting with each other. Thus, it's logical to view a conductor structure system as analogous to a circuit. The concept of an equivalent circuit arises from recognizing that the injector's length behaves like a reactor in response to RF signals, its surfaces act as capacitors, and the ohmic loss in the cavity resembles a resistor in the circuit. In series-connected RLC circuits, the resonant frequency is determined by $\omega = 1/\sqrt{LC}$, where the net impedance is minimized, allowing for the selection of a specific frequency. At this resonant frequency, energy transfer between magnetic energy $LI^{2}/2$ and electric energy $CV^{2}/2$ is most efficient, minimizing apparent energy loss due to phase differences in electromagnetic energy. While LC elements don't consume electromagnetic energy, they do contribute to reducing average energy. Despite this, the similarities between our photoinjector and an LRC circuit, coupled with the need for the cavities in the photoinjector to store as much energy as possible, suggest that an equivalent transmission line can effectively substitute for the photoinjector.


\begin{figure*}
    {
   \subfigure[ Transmission line for a conductor system]{
     \includegraphics[width=9.2 cm, height= 4.8cm]{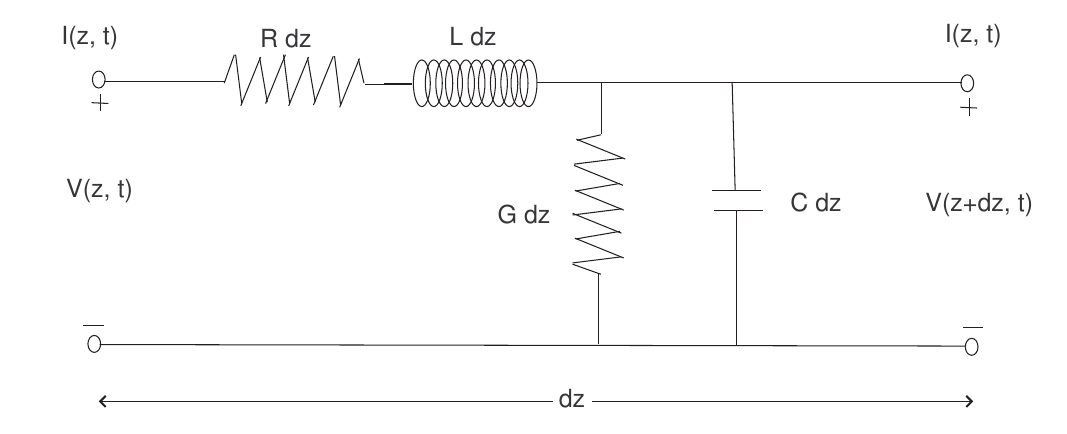}
     \label{plot1a}
    }\hspace{-13 mm}
   \subfigure[Terminated lossless transmission line]{
   \includegraphics[width=9.2 cm,  height= 5.2cm]{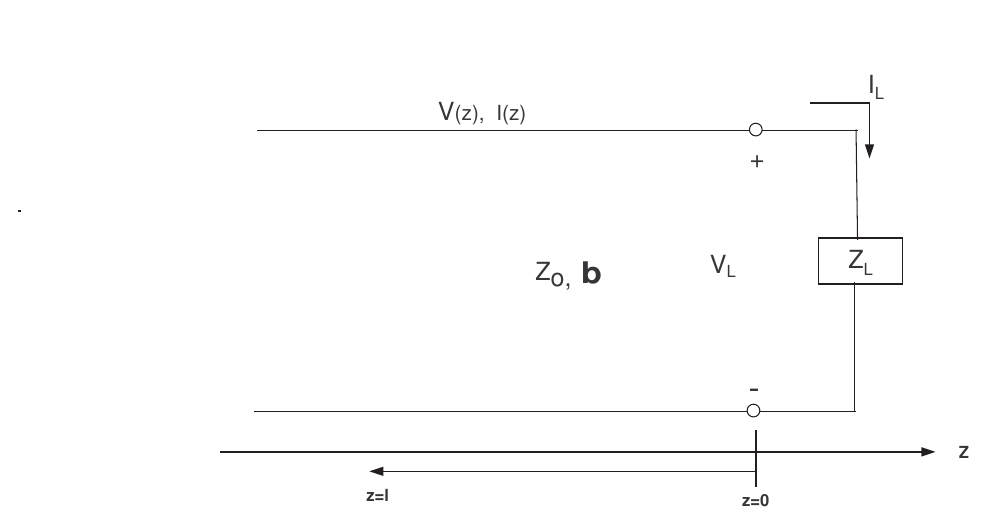}
     \label{plot1b}
   }
}
\caption{}
\end{figure*}

\subsubsection{Transmission line}
Below equations are the relation of voltage and current in a transmission line (Fig.~1)\citep{Microwave_engineering_Pozar}.

\begin{eqnarray}
\frac{dV(z)}{dz} &=& -(R + i\omega L) I(z) \\
\frac{dI(z)}{dz} &=& -(G + i\omega C) V(z)
\end{eqnarray}

The solutions for voltage and current are
\begin{eqnarray}
V(z) &=& V_{0}^{+} e^{-\gamma z} + V_{0}^{-} e^{\gamma z},\\
I(z) &=& \frac{1}{Z_{0}}(V_{0}^{+} e^{-\gamma z} - V_{0}^{-} e^{\gamma z}).
\end{eqnarray}

\begin{eqnarray}
\gamma &=& \alpha + i \beta = \sqrt{(R+i\omega L)(G+i\omega C)},\\
Z_{0}&=& \frac{V^{+}{0}}{I^{+}{0}} = -\frac{V^{-}{0}}{I^{-}{0}}= \frac{R+i\omega L}{\gamma} =
\sqrt{\frac{R+i\omega L}{G+i\omega C}}.
\end{eqnarray}
\\
\textbf{Terminated Lossless Transmission Line}\\\\
General solutions for the circuit are
\begin{eqnarray}
V(z) &=& V_{0}^{+} e^{-i\beta z} + V_{0}^{-} e^{i\beta z},\\
I(z) &=& \frac{1}{Z_{0}}(V_{0}^{+} e^{-i\beta z} - V_{0}^{-} e^{i\beta z}).
\end{eqnarray}
Signals originate from the left part. Some reflect, while others pass through the circuit to the right. Hence, we consider voltage reflection coefficients and the load impedance that the signal encounters in the circuit.\\

The load impedance and voltage reflection coefficient are
\begin{eqnarray}
Z_{L} &=& \frac{V(0)}{I(0)}= \frac{V_{0}^{+} + V_{0}^{-}}{V_{0}^{+} - V_{0}^{-}}Z_{0},\\
\Gamma &=& \frac{V_{0}^{-}}{V_{0}^{+}}=
\frac{Z_{L}-Z_{0}}{Z_{L}+Z_{0}},
\end{eqnarray}
and the total voltage and current waves on the line can be
written like below
\begin{eqnarray}
V(z) &=& V_{0}^{+}[ e^{-i\beta z} + \Gamma e^{i\beta z}]\\
I(z) &=& \frac{V_{0}^{+}}{Z_{0}}( e^{-i\beta z} - \Gamma e^{i\beta z}).
\end{eqnarray}
\noindent The flow of time-averaged power along the line at the
point $z$ and the reflection loss rate are
\begin{eqnarray}
P_{av} = \frac{1}{2}Re[V(z)I(z)^{*}] = \frac{1}{2} \frac{|V_{0}^{+}|^{2}}{Z_{0}}(1-|\Gamma|^{2})\\
RL = -20\, \log \, |\Gamma| \; dB
\end{eqnarray}
\noindent The degree of mismatch $(\Gamma \neq 0)$, called the
\textbf{``standing wave ratio'' (SWR)}, is defined as
\begin{eqnarray}
SWR = \frac{V_{max}}{V_{min}} = \frac{1+|\Gamma |}{1-|\Gamma |}
\end{eqnarray}

The load impedance and voltage reflection coefficient are expressed as follows:
\begin{eqnarray}
Z_{L} &=& \frac{V(0)}{I(0)}= \frac{V_{0}^{+} + V_{0}^{-}}{V_{0}^{+} - V_{0}^{-}}Z_{0},\\
\Gamma &=& \frac{V_{0}^{-}}{V_{0}^{+}}=
\frac{Z_{L}-Z_{0}}{Z_{L}+Z_{0}},
\end{eqnarray}
\noindent while the total voltage and current waves on the line are given by:
\begin{eqnarray}
V(z) &=& V_{0}^{+}[ e^{-i\beta z} + \Gamma e^{i\beta z}],\\
I(z) &=& \frac{V_{0}^{+}}{Z_{0}}( e^{-i\beta z} - \Gamma e^{i\beta z}).
\end{eqnarray}
\noindent The flow of time-averaged power along the line at point $z$ and the reflection loss rate are computed as:
\begin{eqnarray}
P_{av} &=& \frac{1}{2}Re[V(z)I(z)^{*}] = \frac{1}{2} \frac{|V_{0}^{+}|^{2}}{Z_{0}}(1-|\Gamma|^{2}),\\
RL &=& -20, \log , |\Gamma| ; dB.
\end{eqnarray}
\noindent Additionally, the degree of mismatch $(\Gamma \neq 0)$, known as the standing wave ratio(SWR), is defined by
\begin{eqnarray}
SWR &=& \frac{V_{max}}{V_{min}} = \frac{1+|\Gamma |}{1-|\Gamma |}.
\end{eqnarray}

\subsubsection{Quarter Wave Transformer}
This concept is applicable to understanding standing waves in a mismatched line or impedance matching circuit structure\citep{Microwave_engineering_Pozar}. For instance, consider a circuit where the load resistance $R_{L}$ and characteristic impedance $Z_{0}$ are connected with a lossless piece of transmission line. The characteristic impedance of this transmission line is $Z_{1}$, and its length is $\lambda /4$. To ensure that the reflection coefficient $\Gamma$ is zero, we need to determine the condition of $Z_{1}$. The impedance perceived by the signal in the circuit is determined by:

\begin{eqnarray}
Z_{in}=Z_{1}\frac{R_{L}+ i Z_{1}\tan \beta \ell}{Z_{1}+i R_{L} \tan \beta \ell}.
\end{eqnarray}
For the case of $\beta \ell = (2\pi /\lambda)(\lambda/4)$, the impedance is
\begin{eqnarray}
Z_{in}=\frac{Z_{1}^{2}}{R_{L}}.
\end{eqnarray}
To make $\Gamma = 0$, $Z_{in}$ should be the same as $Z_{0}$ (Eqs. 3.10)and then $Z_{1}$ is
\begin{eqnarray}
Z_{1}=\sqrt{Z_{0} R_{L}}.
\end{eqnarray}
This result indicates that the reflection coefficient $\Gamma$ can be minimized with the appropriate impedance $Z_{1}$ and a length of $\lambda / 4$. This straightforward theory will be applied in connecting the waveguide and full-cell cavity to enhance the efficiency of energy transfer.

\subsubsection{Scattering Matrix}
To quantify the energy entering the injector, we utilize the S-parameter matrix. This matrix elucidates the relationship between the phase of incident and outgoing voltages at a port. The magnitude and phase difference of the S-matrix denote the ratio of output to input signal voltage and their phase discrepancies.

\begin{align}
\begin{pmatrix}
V^{-}_{1} \\  V^{-}_{2} \\ \vdots \\ V^{-}_{n}
\end{pmatrix} &=
\begin{pmatrix}
S_{11} & S_{12} & \cdots & S_{1n} \\
S_{21} & S_{22} & \cdots & S_{2n} \\
\vdots & \vdots & \ddots & \vdots \\
S_{n1} & S_{n2} & \cdots & S_{nn}
\end{pmatrix}
\begin{pmatrix}
V^{+}_{1} \\  V^{+}_{2} \\ \vdots \\ V^{+}_{n}
\end{pmatrix}
\end{align}
\begin{eqnarray}
S_{ij} = \frac{V^{-}_{i}}{V^{+}_{j}} \bigg |_{V^{+}_{k}=0 \; for \; k \neq j}
\end{eqnarray}

\noindent Here, $S_{ii}$ represents the reflection coefficient observed at the $ith$ port.

\subsection{RF Coupling}
We will create an equivalent network for a 1.6-cell photoinjector and solve it using the small aperture approximation [9], which accounts for the coupling between the rectangular waveguide and two cavities. Following an introduction to the electric and magnetic normalized field vectors $\vec{F}$ and dipole vector $\vec{d}$, we will review the theory of the small aperture approximation. Subsequently, we will construct the equivalent circuit of the 1.6-cell photoinjector based on the results obtained. The scattering matrix, derived from the reaction tensor, will be employed. While the methodologies in this section draw from Lin et al.\citep{1995PhDT........39L, 1996pac..conf..954L}, Shapiro \citep{2001PhRvS...4d2001S}, and Brown et al. \cite{1999NIMPA.425..441B, 1999AIPC..472..861B}, certain modifications and additions have been made to accommodate the structure of the photoinjector.

\begin{figure*}
    {
   \subfigure[]{
     \includegraphics[width=7.2 cm, height= 7cm]{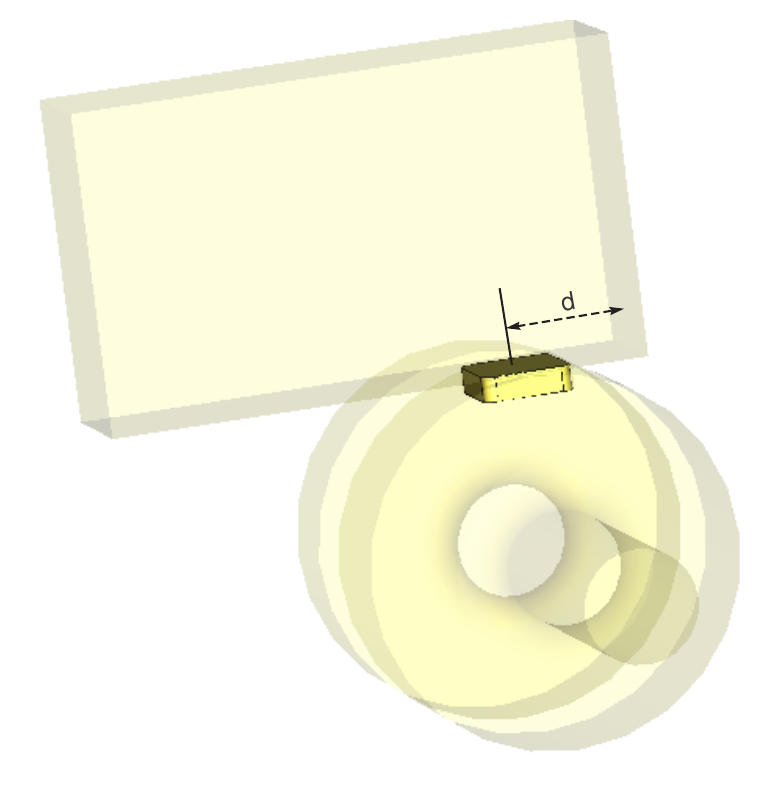}
     \label{plot2a}
    }\hspace{-0 mm}
   \subfigure[]{
   \includegraphics[width=10.2 cm, height= 8cm]{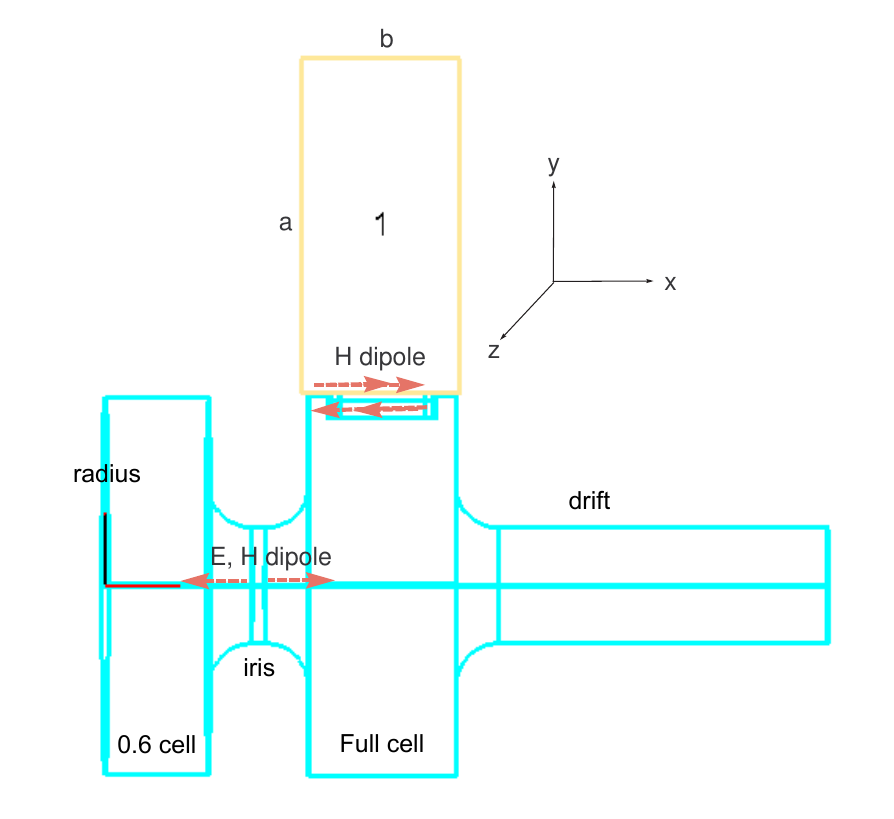}
     \label{plot2b}
   }
}
\caption{d $\rightarrow \lambda/4$}
\end{figure*}

\subsubsection{Small Aperture Approximation}
The concept of waves entering a cavity through small apertures is elucidated by Bethe's theory \citep{1944PhRv...66..163B}. According to this theory, any changes in electric charge and magnetic current density near these apertures act as radiation sources. Specifically, the electric field perpendicular to the aperture creates an electric dipole, with its magnitude proportional to the strength of the electric field. Similarly, the magnetic field parallel to the aperture generates a magnetic dipole, with its magnitude proportional to the magnetic field strength. Both the electric and magnetic dipoles oscillate at the frequency of the incident electromagnetic fields, thereby radiating wave energy and acting as new wave sources. The coefficient tensors, which are proportional to the induced electromagnetic dipoles, are respectively termed the electrical polarizability ($\alpha_{e}$) and magnetic polarizability ($\alpha_{m}$). These coefficients are contingent upon the geometric shape and size of the aperture and are determined using the continuity of potential and boundary conditions \cite{1998clel.book.....J}.\\

If $r_{0}$ represents the radius of a circular aperture and $d$ denotes the length of a rectangular aperture, the expressions for the dipoles are as follows:
\begin{eqnarray}
\bf{P}_{e} = \epsilon_{0} \,\alpha_{e}\, E_{n} \,\delta(r-r_{0})\hat{n},\quad \bf{P}_{m} = -\alpha_{m} \, H_{t}\,\delta(r-r_{0})\hat{t}
\end{eqnarray}
$$\begin{tabular} { l| l l  } \hline Aperture shape &   $\alpha_{e}$  &  $\alpha_{m}$ \\ \hline
Round hole & $\frac{2 r_{0}^{3}}{3}$   & $\frac{4 r_{0}^{3}}{3}$\\
Rectangular slot & $\frac{\pi \ell d^{2}}{16}$   & $\frac{\pi \ell d^{2}}{16}$\\ \hline
\end{tabular}$$
Near the apertures, Maxwell equations are written as
\begin{eqnarray}
\nabla \times \bf{E} &=& -i \omega \mu_{0}(\bf{H} + \bf{P_{m}}) - \bf{M}, \\
\nabla \times \bf{H} &=& i \omega (\epsilon_{0}\bf{E} + \bf{P_{e}}) +\bf{J}.
\end{eqnarray}
Now, we can describe the $E$ and $H$ fields formed by the sources using the mode excitation method.\\

For the frequency of 2.856 GHz and the geometry of the rectangular waveguide, the $TE_{10}$ mode propagates through the rectangular waveguide. The dipole and the  polarizabilities are represented as follows:
\begin{eqnarray}
\tensor{\alpha}_{e}=\alpha_{ex}\; \hat{x}\hat{x},\,\,\tensor{\alpha}_{m}=\alpha_{my}\; \hat{y}\hat{y} + \alpha_{mz} \; \hat{z}\hat{z}.
\end{eqnarray}
We define a field vector $\vec{F}$ and a dipole vector
$\vec{d}$ as
\begin{eqnarray}
\vec{F} &=& \frac{E_{x}}{\eta}\hat{x} + H_{y} \hat{y} + H_{z} \hat{z},\\
\vec{d} &=& P_{mz}\hat{z},\\
\eta &=& \sqrt{\frac{\mu_{0}}{\epsilon_{0}}},
\end{eqnarray}
where the dipole vector can be written as
\begin{eqnarray}
\vec{d} &=& \tensor{\alpha} \cdot \vec{F}^{guide} = \alpha_{mz} H_{z}\, \hat{z} =
P_{mz}\hat{z}.
\end{eqnarray}
The net fields $\mathbf{E}$ and $\mathbf{H}$ are
\begin{eqnarray}
\vec{F}_{net} &=& \vec{F}^{inc}+ \vec{F}^{g} + \vec{F}^{c}\,\,\,(\mathrm{g:\, guide, c:\, cavity})
\end{eqnarray}

\subsubsection{Reaction Tensors}
The field vector has a relation with the previously defined dipole vector, but their connection can be further elucidated using a reaction tensor $\chi$.
\begin{eqnarray}
\vec{F}^{g} = \tensor{\chi}^{g} \cdot \vec{d}, \quad  \vec{F}^{c} = \tensor{\chi}^{c} \cdot \vec{d}
\end{eqnarray}
Then,
\begin{eqnarray}
\vec{d} = \tensor{\alpha} \cdot \vec{F}^{inc}+ \tensor{\alpha} \cdot (\tensor{\chi}^{g} + \tensor{\chi}^{c}) \cdot \vec{d} \quad
\Rightarrow \quad \vec{d} = [\tensor{\delta} - \tensor{\alpha} \cdot (\tensor{\chi}^{g} + \tensor{\chi}^{c})]^{-1}\cdot \tensor{\alpha} \cdot \vec{F}^{inc}
\end{eqnarray}
The process entails a recursive method, wherein these outcomes will be employed to differentiate the dipole vector from the field vector, subsequently utilized in computing the S matrix. The primary objective of this section is to derive the reaction tensor.

\subsubsection{Mode Excitation from Electric and Magnetic Dipole Source}
The electric or magnetic field in a waveguide can be expanded with other normalized modes.
\begin{eqnarray}
E^{+}&=&\sum_{n}A_{n}^{+} E_{n}^{+} = \sum_{n} A_{n}^{+}(\overline{e}_{n} + \hat{z}e_{zn})\,e^{-i\beta_{n}
z},\,
z> z_{2}\\
H^{+}&=&\sum_{n}A_{n}^{+} H_{n}^{+} = \sum_{n} A_{n}^{+}(\overline{h}_{n} + \hat{z}h_{zn})\,e^{-i\beta_{n}
z},\,
z> z_{2}\\
E^{-}&=&\sum_{n}A_{n}^{-} E_{n}^{-} = \sum_{n} A_{n}^{-}(\overline{e}_{n} - \hat{z}e_{zn})\,e^{i\beta_{n} z},\,
z< z_{1}\\
H^{-}&=&\sum_{n}A_{n}^{-} H_{n}^{-} = \sum_{n} A_{n}^{-}(-\overline{h}_{n} + \hat{z}h_{zn})\,e^{i\beta_{n}
z},\, z< z_{1}
\end{eqnarray}
The coefficients are \citep{Microwave_engineering_Pozar}
\begin{eqnarray}
A_{n}^{+} &=&  \frac{1}{P_{n}}\int_{V}  [-(\overline{e}_{n} - \hat{z}e_{zn})\cdot J +(-\overline{h}_{n} +
\hat{z}h_{zn}) \cdot M] e^{i\beta_{n} z} \; dv \cr &=&\frac{1}{P_{n}}\,\int_{V}  [-E^{-}_{n}\cdot J
+H_{n}^{-} \cdot M] e^{i\beta_{n} z}  dv \label{coefficient_A_plus} \\
A_{n}^{-} &=&  \frac{1}{P_{n}}\int_{V}  [-(\overline{e}_{n} + \hat{z}e_{zn})\cdot J +(\overline{h}_{n} +
\hat{z}h_{zn}) \cdot M] e^{i\beta_{n} z}  dv \cr
&=&\frac{1}{P_{n}}\,\int_{V}  [-E^{+}_{n}\cdot J + H_{n}^{+} \cdot M] e^{i\beta_{n} z} dv \quad (P_{n} = 2\int_{S_{0}} \overline{e}_{n} \times \overline{h}_{n}\cdot \hat{z} \, dS) \label{coefficient_A_minus}
\end{eqnarray}
Considering that the electric and magnetic dipoles function as electric and magnetic current sources at the aperture located at $\delta(x)\delta(y)\delta(z)$, we can express the current sources as follows:
\begin{eqnarray}
J &=& i\omega P_{e} \delta(x)\delta(y)\delta(z)\\
M &=& i\omega \mu_{0} P_{m} \delta(x)\delta(y)\delta(z)
\end{eqnarray}

\begin{figure*}
    {
   \subfigure[]{
     \includegraphics[width=9.3cm]{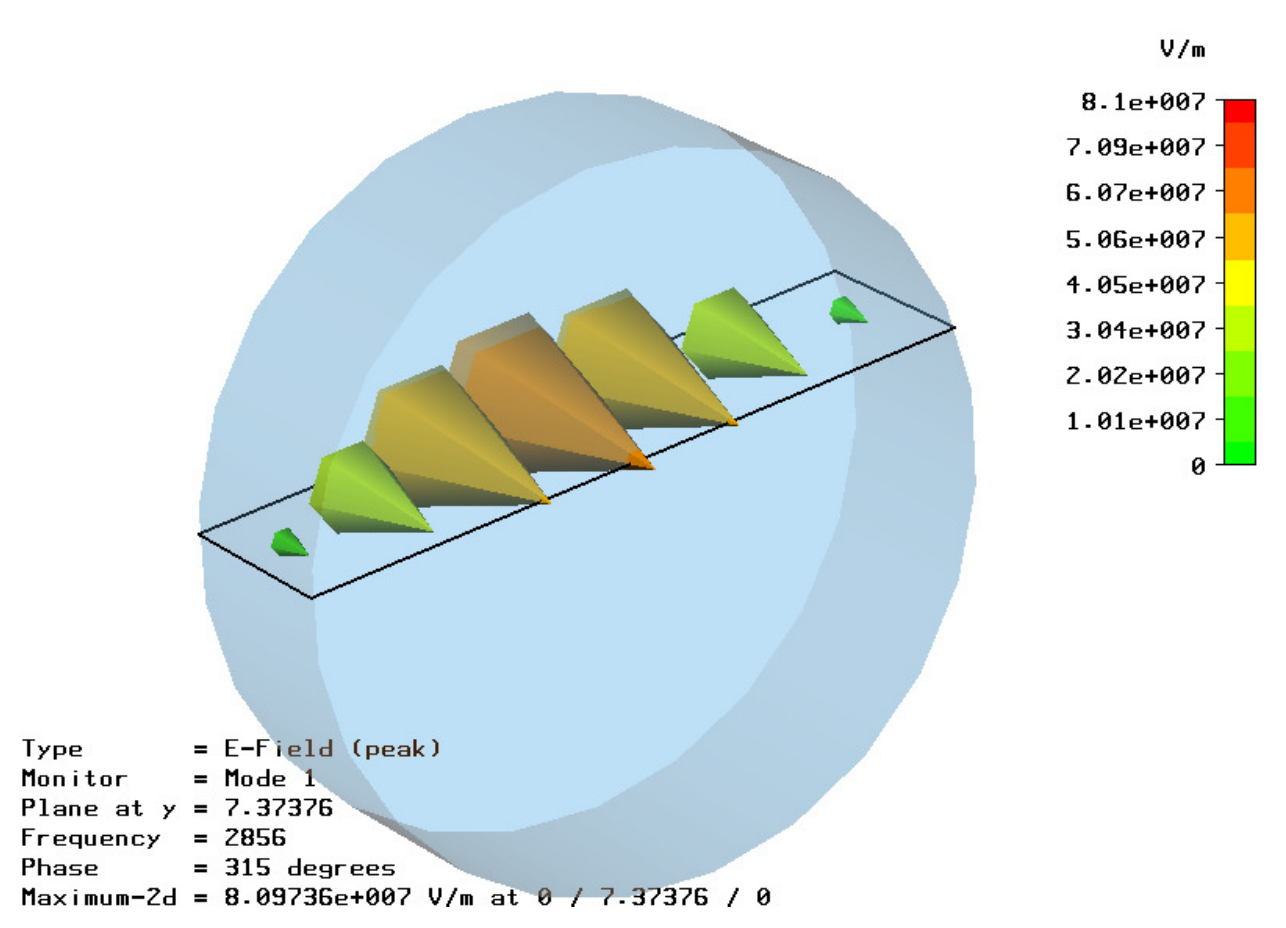}
     \label{plot3a}
    }\hspace{-0 mm}
   \subfigure[]{
   \includegraphics[width=7.5cm]{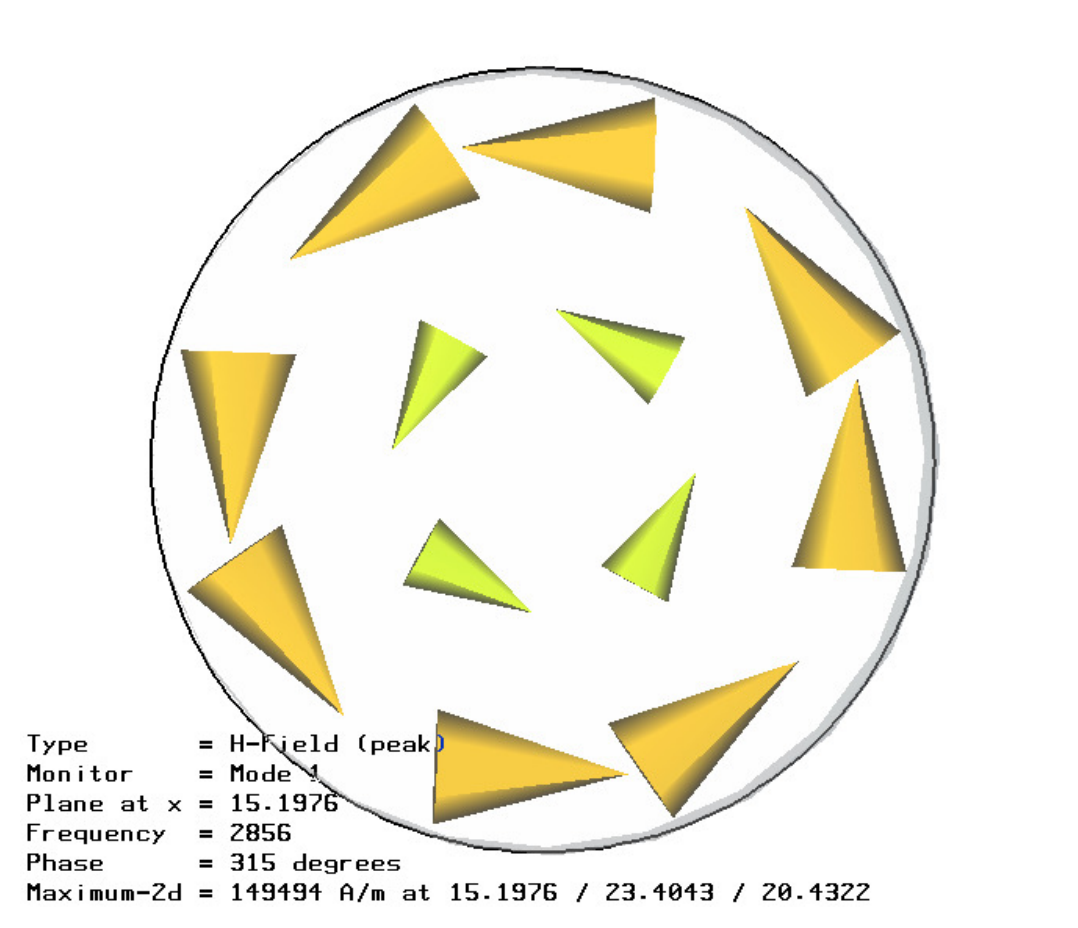}
     \label{plot3b}
   }
}
\caption{The E field is strongest on the axis and weakens as the position moves away from the axis. Conversely, the H field is strongest at the surface and gradually weakens as the position approaches the axis. This mode demonstrates the property of minimizing energy loss, as the Q factor becomes large.}
\end{figure*}

\section{Theoretical Analysis: Construction of the equivalent circuit}
$TE_{10}$ mode propagating into a rectangular waveguide arrives at a hole aperture and radiates into the cavity. Using mathematical symbols and physical concepts, we will calculate impedances, with which we construct a corresponding equivalent circuit. The results, such as currents and resistance from solving the circuit, provide valuable insights into their physical properties. In the model of \citep{1996pac..conf..954L}, the wave enters through two apertures located at the tops of both the half and full cell. The dipoles formed at the two apertures include $H_{x}$, $E_{y}$, and $H_{z}$ components in the $TE_{10}$ mode, representing electric and magnetic dipoles. However, since the electric dipole is prone to causing voltage breakdown, it is desirable to exclude the electric component when forming an equivalent dipole. To do so, there are two choices: upright attachment of a rectangular waveguide vertically or attaching the rectangular one. For the latter case, only $H_{z}$, which is parallel to the plane of the aperture, forms a magnetic dipole. Besides reducing the possibility of voltage breakdown by the electric dipole, the calculation becomes much simpler. Moreover, the result of the analytical calculation does not have nonlinear terms. Lastly, we explore how the energy reflection coefficient can be diminished in this model.\\

\subsection{Waveguide and Cavity Reaction Tensor}
The normalized wave functions for a rectangular waveguide are
\begin{eqnarray}
e_{1}^{\pm}\hat{x} &=&  E_{0}\,\sin \frac{\pi y}{a}\, \hat{x},\\
h_{1}^{\pm}\hat{y} &=& \mp H_{0}^{'}\,\sin \frac{\pi y}{a} \, \hat{y}, \quad (\int_{S_{apt}}
e_{1}^{(\pm)} \times h_{1}^{(\pm)} \cdot \hat{z} dS = 1)\\
h_{1}^{\pm}\hat{z}&=& H_{0}\, \cos \frac{\pi x}{a}\, \hat{z}.
\end{eqnarray}
We can calculate the coefficients of scattered waves [Eq.~(\ref{coefficient_A_plus}), (\ref{coefficient_A_minus})].
\begin{eqnarray}
A_{1}^{+} = A_{1}^{-}= \frac{i\omega \mu_{}}{2}h_{1z}^{\mp}P_{mz}
\end{eqnarray}
In principle, more coefficients can be expanded for an exact result, but it is experimentally known that the first term ($n = 1$) gives reasonable values in many cases. Then, the above coefficients can be factored as follows:
\begin{eqnarray}
A_{1}^{+} &=& \frac{i\omega \mu_{}}{2} \left( \begin{matrix} \sqrt{\epsilon / \mu}\,e_{1x}^{-} & h_{1y}^{-} & h_{1z}^{-} \end{matrix} \right)  \left( \begin{matrix} 0 \\ 0 \\ P_{mz} \end{matrix} \right) \equiv  \frac{i\omega \mu_{}}{2} \vec{f}^{-}\cdot \vec{d}, \\
A_{1}^{-} &=& \frac{i\omega \mu_{}}{2} \left( \begin{matrix} \sqrt{\epsilon / \mu}\,e_{1x}^{+} & h_{1y}^{+} & h_{1z}^{+} \end{matrix} \right) \left( \begin{matrix} 0 \\ 0 \\ P_{mz} \end{matrix} \right)  \equiv  \frac{i\omega \mu_{}}{2} \vec{f}^{+}\cdot \vec{d}.
\end{eqnarray}
Then, the field vectors are
\begin{eqnarray}
\vec{F}^{guide} = \tensor{\chi}_{g}\cdot \vec{d} = \frac{1}{2}(A^{+}\vec{f}^{+} + A^{-}\vec{f}^{-})
= \frac{i\omega \mu}{4}[(\vec{f}^{-}\cdot \vec{d})\,\vec{f}^{+} + (\vec{f}^{+}\cdot \vec{d})\,\vec{f}^{-}]
\end{eqnarray}
Thus,
\begin{eqnarray}
\chi_{g} = \frac{i\omega \mu}{4} \left(
\begin{matrix}
0 & 0 & 0 \cr
0 & 0 & 0 \cr
0 & 0 & 2f_{z}^{2}
\end{matrix}
\right).
\end{eqnarray}
The dipole formed by the $TE_{10}$ mode near the aperture hole oscillates due to the oscillation of the incident $TE_{10}$ mode. This oscillation radiates new electromagnetic waves into the injector, creating the $TM_{010}$ mode in the cavity. Referring to Fig.~3, we know that only the $H_{\theta}$ field exists on the surface, and its direction coincides with $\hat{z}$. This property simplifies the description of the wave mode. According to Bethe's theory, only the $H_{t}$ and $E_{n}$ components generate the equivalent dipole at a hole, meaning that the components of other fields except $H_{\theta}$ do not generate equivalent dipoles.\\

\begin{eqnarray}
H_{\theta} \Rightarrow H_{z} = i \frac{E_{0}}{\mu c}J_{1}(\frac{p_{01}\; r}{R_{cav}})
\end{eqnarray}
And we assume that the frequency of the full cavity is the same as that of the dipole, as will become clear later. From the Maxwell equations,
\begin{eqnarray}
\nabla \times H &=& i \omega_{c} \epsilon E \\
\nabla \times E &=& -i \omega_{c} \mu H -i \omega \mu P_{m} \delta(x-\frac{b}{2})\delta(y)\delta(z).
\end{eqnarray}
\noindent With $\nabla \times (\nabla \times H)$,
\begin{eqnarray}
-\nabla^{2} H &=& k^{2} H \simeq  \omega_{c}^{2} \mu \epsilon H + \omega^{2} \mu \epsilon P_{m}
\delta(x-\frac{b}{2})\delta(y)\delta(z)
\end{eqnarray}
After integration, we obtain,
\begin{eqnarray}
\pi R_{c2}^{2} \ell_{c2} H_{z} =
\frac{\omega^{2}}{\omega^{2}-\omega_{c2}^{2}} P_{m}
\end{eqnarray}
With $\omega_{c2} = \sqrt{1/L_{2}C_{2}}$, the above result can be written as
\begin{eqnarray}
H_{z} &=& \frac{1}{i\omega C_{2} + \frac{1}{i\omega L_{2}}}\; \frac{i\omega C_{2}}{\pi R_{c_{2}}^{2}
\ell_{c_{2}}}\; P_{m},
\end{eqnarray}
where
\begin{eqnarray}
C_{2}= \frac{4W_{e}}{V^{2}} = \frac{\int \epsilon |E|^{2} \, dV}{(E_{0}\ell_{c_{2}})^{2}}= \frac{\pi \epsilon
R^{2}_{c_{2}}|J_{0}({p_{01}})|^{2}}{\ell_{c_{2}}}.
\end{eqnarray}
The reaction tensor for the cavity is,
\begin{eqnarray}
\chi_{c_{2}} &=&
\left(
\begin{matrix}
0 & 0 & 0 \\
0 & 0 & 0 \\
0 & 0 & \frac{1}{i\omega C_{2} + \frac{1}{i\omega L_{2}}} \cdot \frac{i\omega C_{2}}{\pi R_{c_{2}}^{2} \ell_{c_{2}}}
\end{matrix}
\right)
\end{eqnarray}
Defining $F^{inc} = A_{0} f^{+}$, we get the final dipole vector representation as follows:
\begin{eqnarray}
\vec{d} &=& [I-\alpha \cdot (\chi_{g}+ \chi_{c_{2}})]^{-1}\cdot \alpha \cdot F_{inc}\nonumber \\
&=& A_{0} \left[\left(
\begin{matrix}
1&0&0 \\
0 & 1 & 0 \\
0 & 0 &  1
\end{matrix}
\right)-
\left( \begin{matrix}
\alpha_{x}&0&0 \\
0 & \alpha_{y} & 0 \\
0 & 0 & \alpha_{z}
\end{matrix}
\right)\cdot
\left( \begin{matrix}
0 & 0 & 0 \\
0 & 0 & 0 \\
0 & 0 & \frac{i \omega \mu f_{z}^{2}}{2} + \frac{1}{i\omega C_{2} + \frac{1}{i\omega L_{2}}}\; \frac{i\omega C_{2}}{\pi R_{c_{2}}^{2} \ell_{c_{2}}}
\end{matrix}
\right) \right]^{-1}\cdot \alpha \cdot f^{+}
\end{eqnarray}

\subsection{Equivalent circuit for the waveguide and cavity}
\subsubsection{Equivalent circuit for the waveguide}
If we observe the waves near the aperture, they resemble the signals in a two-port transmission line system. The S matrix in the system illustrates the relationship between incoming and outgoing voltage. Therefore, if we transform the photoinjector and the waves into a corresponding transmission line, the S matrix provides us with physically interpretable information on the waves and their behaviors. In transmission line theory, the S matrix is defined as follows, and impedance is simply the inverse of the S matrix (considering the symmetry, $S_{12}=S_{21}, S_{11}=S_{22}$),
\begin{eqnarray}
[S] &=& ([Z]-[I])^{-1}([Z]+[I])\\
\cr Z &=& \frac{1}{(1-S_{11})^{2}- S_{12}^{2}}  \left( \begin{matrix}
1-S_{11}^{2}+S_{12}^{2} &
2S_{21}\cr 2S_{21} &
1-S_{11}^{2}+S_{12}^{2}\cr
\end{matrix}
\right)
\end{eqnarray}
The length between the aperture and the end wall of the waveguide [Fig.~2(a)] is a quarter wavelength long, denoted as \( d = \lambda/4 \). Consequently, the phase difference between the incident and reflected waves is \( \lambda \), implying there is no actual phase difference between them. In transmission line theory, the end wall of the waveguide, acting as the second port, is considered an open circuit. This effectively reduces the two-port system to a one-port network with only the \( Z_{11} \) impedance. With only the \( Z_{11} \) impedance present, it becomes straightforward to draw the equivalent circuit and solve it because \( Z_{11} \) is simply an expression of parallel and series connections of elements like resistors, etc. Therefore, the primary goal is to obtain \( S_{11} \) and \( S_{21} \). This can be achieved from the definition of the S matrix and the field vector.\\
\begin{eqnarray}
S_{11} &=& \frac{V_{1}^{-}}{V_{1}^{+}}\bigg|_{V_{2}^{+}=0} \Rightarrow  \frac{A^{-}}{A_{0}}= \frac{i\omega
\mu}{2} f^{+}\cdot d \cr &=& \frac{i\omega \mu}{2} f^{+}\cdot  [I-\alpha \cdot (\chi_{g}+ \chi_{c})]^{-1}\cdot \alpha \cdot f^{+},\\
S_{21} &=& \frac{V_{2}^{-}}{V_{1}^{+}}\bigg|_{V_{2}^{+}=0}\Rightarrow  1 + \frac{A^{+}}{A_{0}} = 1 +
\frac{i\omega \mu}{2} f^{-}\cdot d,
\end{eqnarray}
where $A_{0}$ serves as a normalization factor. The impedance is $Z_{11}$ is
calculated exactly as follows:
\begin{eqnarray}
Z_{11} &=& \frac{[\chi_{c2}]_{33}}{i\mu \omega f_{z}^{2}}
+i\frac{1}{\mu \omega f_{z}^{2} \alpha_{z}}, \label{impedance_Z11}\\
\frac{[\chi_{c2}]_{33}}{i \mu \omega f_{z}^{2}} &=&
\frac{C_{2}}{(i\omega C_{2}  + \frac{1}{i\omega L_{2}})\mu \pi
R_{2}^{2}\ell_{2} f_{z}^{2}} \equiv \frac{n_{2}^{2}}{i\omega C_{2} +
\frac{1}{i\omega L_{2}}}.
\label{impedance_Z11_2}
\end{eqnarray}
The first term of Eq.~(\ref{impedance_Z11}) is related to energy transfer, and the second one represents impedance or susceptance $iX_{3z}$. And, `$n_{2}$' indicates a transformer.  The magnetic dipole delivers energy through the aperture and works as a kind of impedance as well.

\begin{figure*}
  \includegraphics[height = 5.6 cm, width=18 cm]{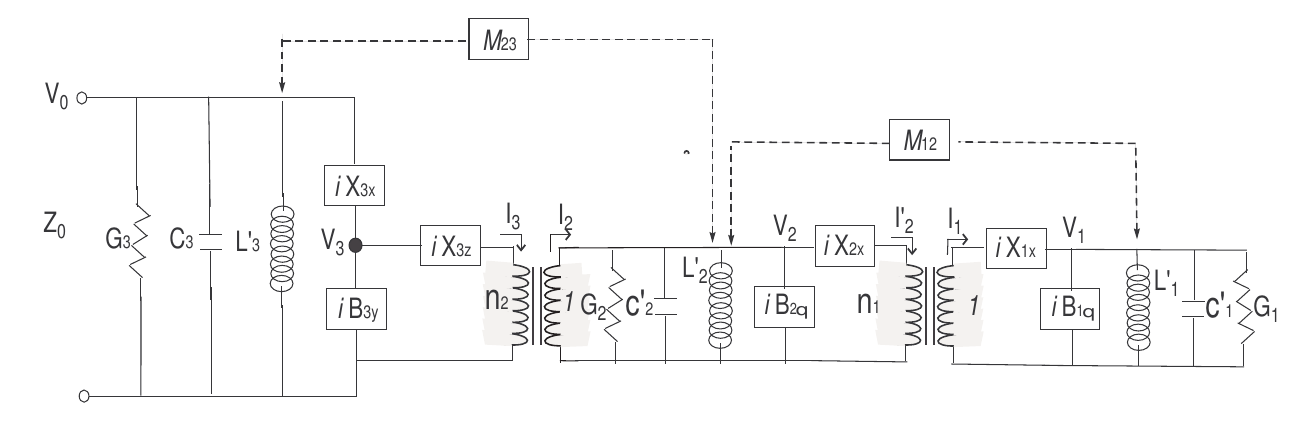}
     \label{plot4}
\caption{Equivalent circuit for the photoinjector}
\end{figure*}

\begin{figure*}
  \includegraphics[width=14cm,  height= 14cm]{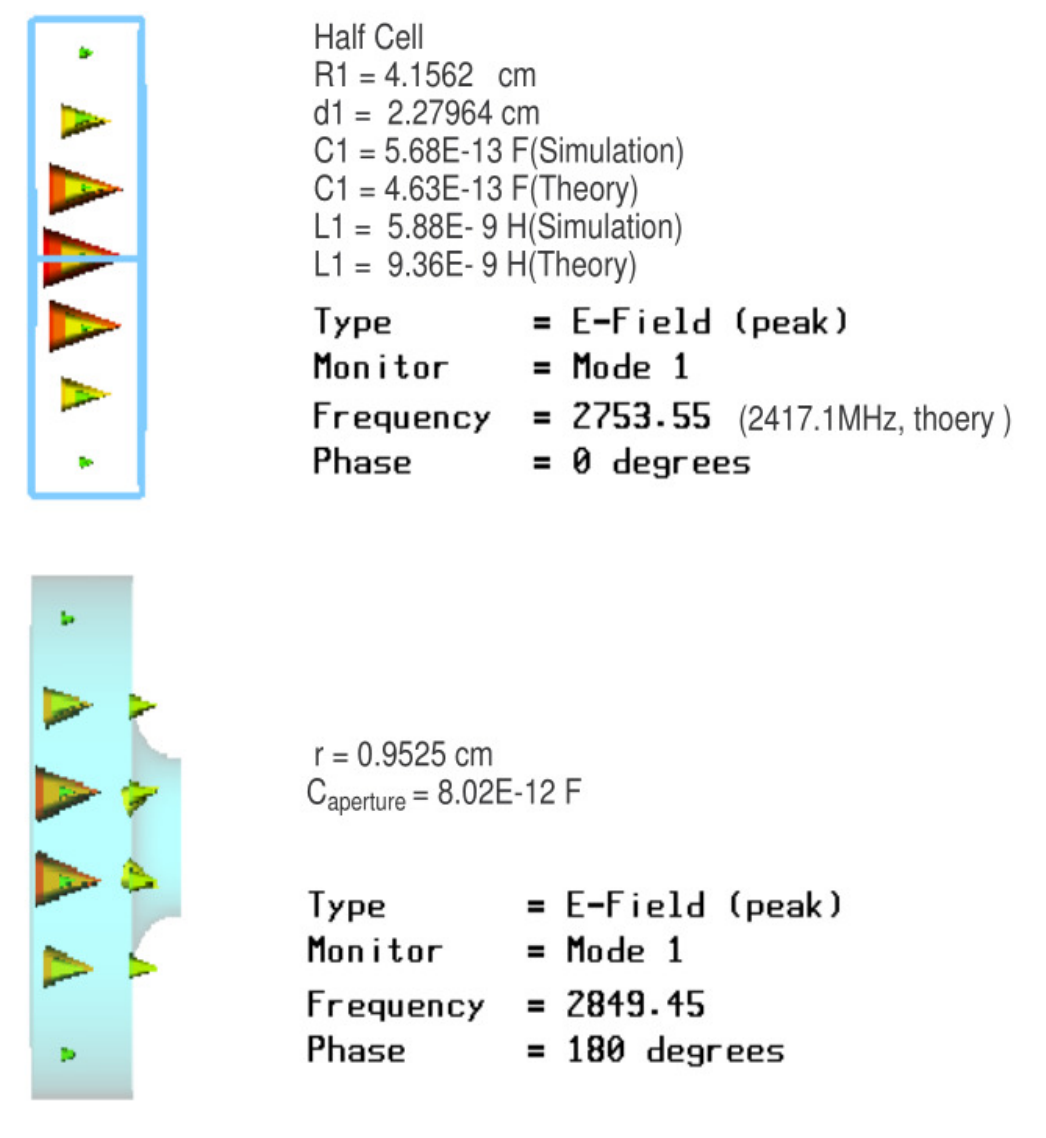}
     \label{plot5}
\caption{The resonant frequency of the pillbox has increased from 2753.54 MHz to 2849.45 MHz due to the strong electric field magnitude near the iris. This change exemplifies a typical case of outward perturbation.}
\end{figure*}

\begin{figure*}
  \includegraphics[width=14cm,  height= 14cm]{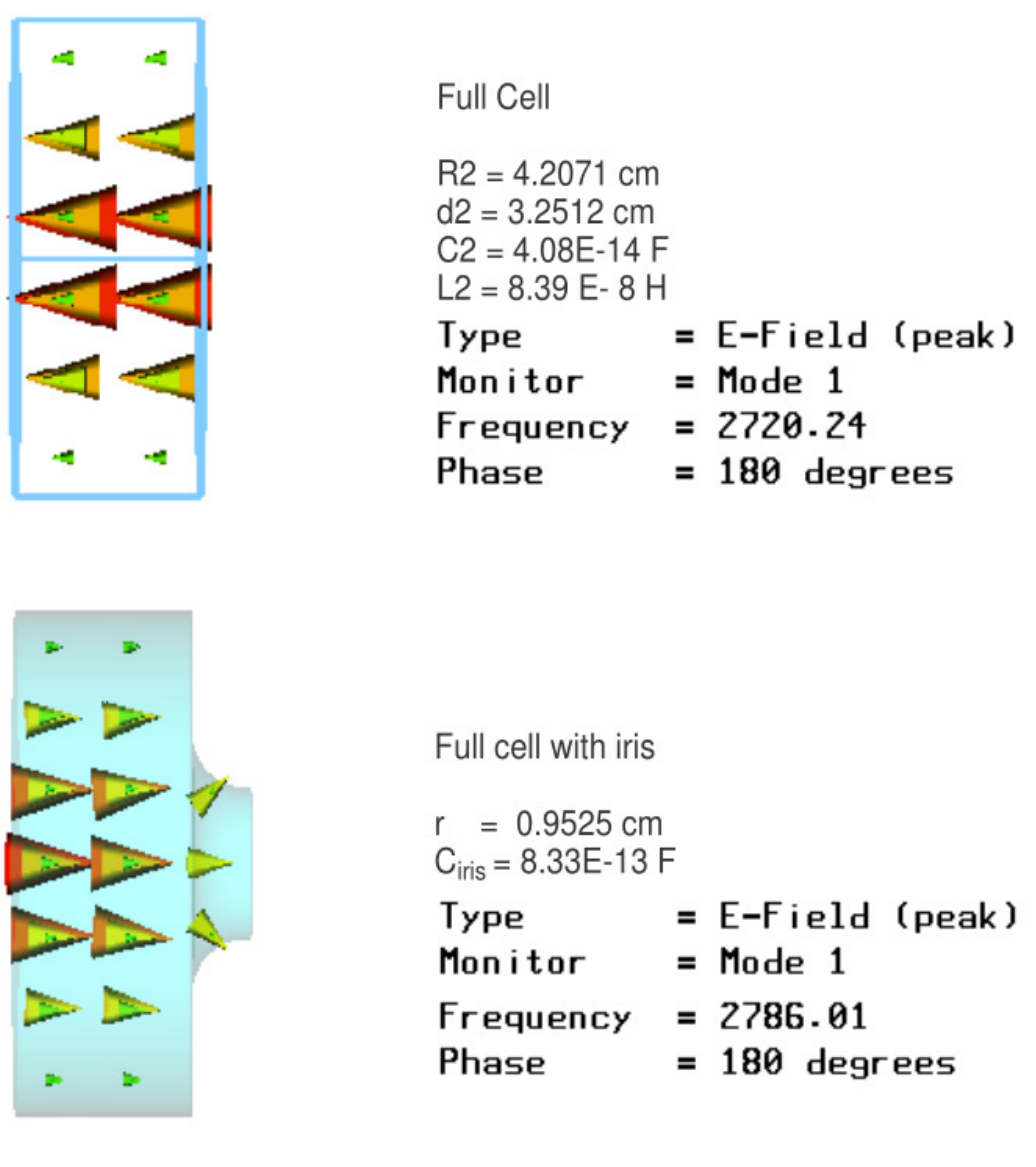}
     \label{plot6}
\caption{The results for the full cell are analogous to the half-cell scenario. However, due to the enlarged volume, the frequency has decreased. The presence of the iris near the axis contributes to an increase in the resonant frequency.}
\end{figure*}

\begin{figure*}
  \includegraphics[width=14cm]{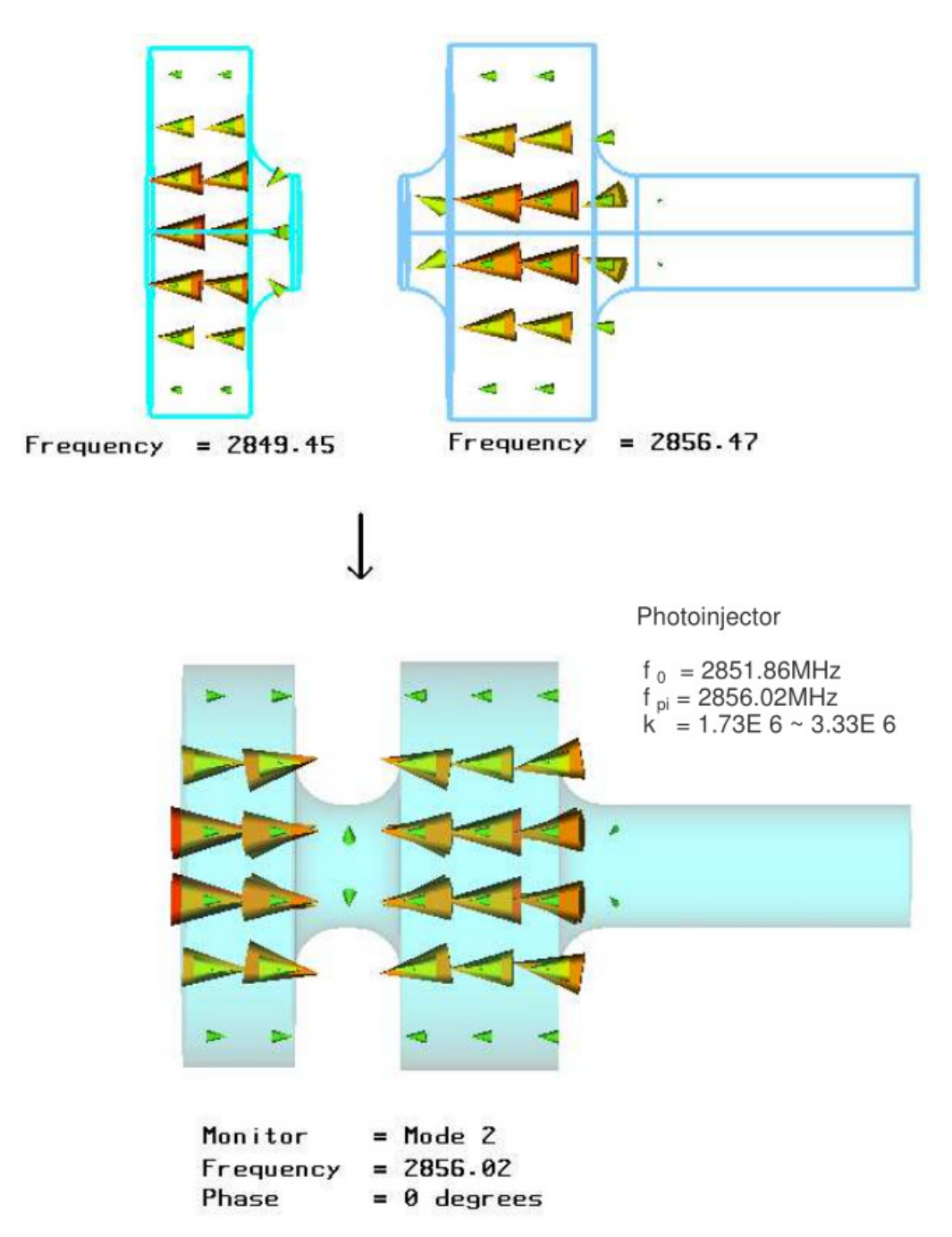}
     \label{plot7}
\caption{Attaching the drift enlarges the E-field area, resulting in an increase in the resonant frequency by approximately 77 MHz. It's noteworthy that the E-field appears to exist in the drift region, despite its cut-off frequency being higher than that of the mode. The lower picture illustrates how the aperture, where the H-field is strong, decreases the frequency. The field observed in the rectangular aperture is H.}
\end{figure*}

\begin{figure*}
  \includegraphics[width=16 cm]{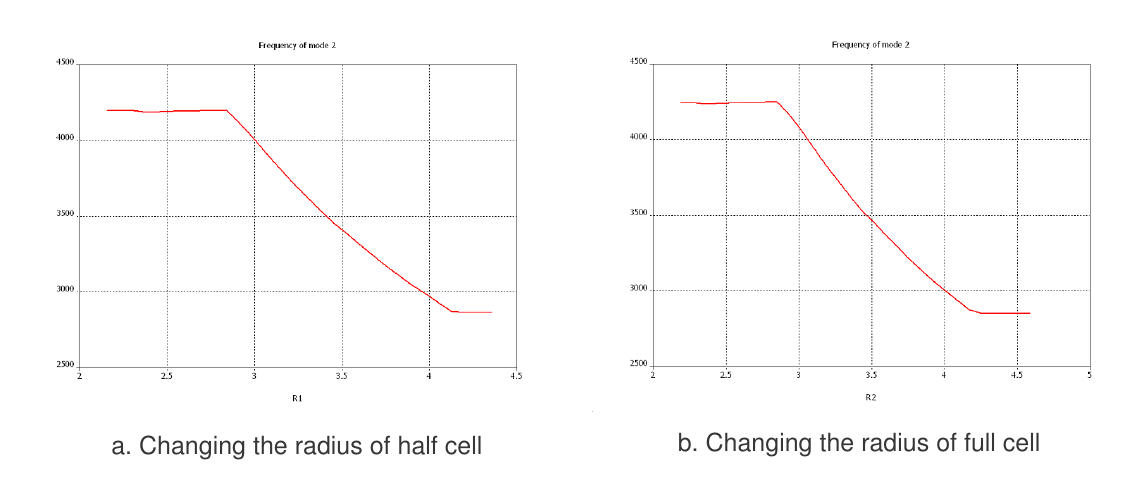}
     \label{plot8}
\caption{Attaching two cavities creates a complete photoinjector. The resonant frequency of the half cell is 2849.45 MHz, and that of the full cell is 2856.47 MHz. This results in a photoinjector operating at 2856.02 MHz in $\pi$ mode. The coupling constant ranges from $1.73 \times 10^{6}$ to $3.33 \times 10^{6}$ depending on the calculation. The discrepancy in these values is attributed to numerical errors such as round-off.}
\end{figure*}

\begin{figure*}
  \includegraphics[width=14cm]{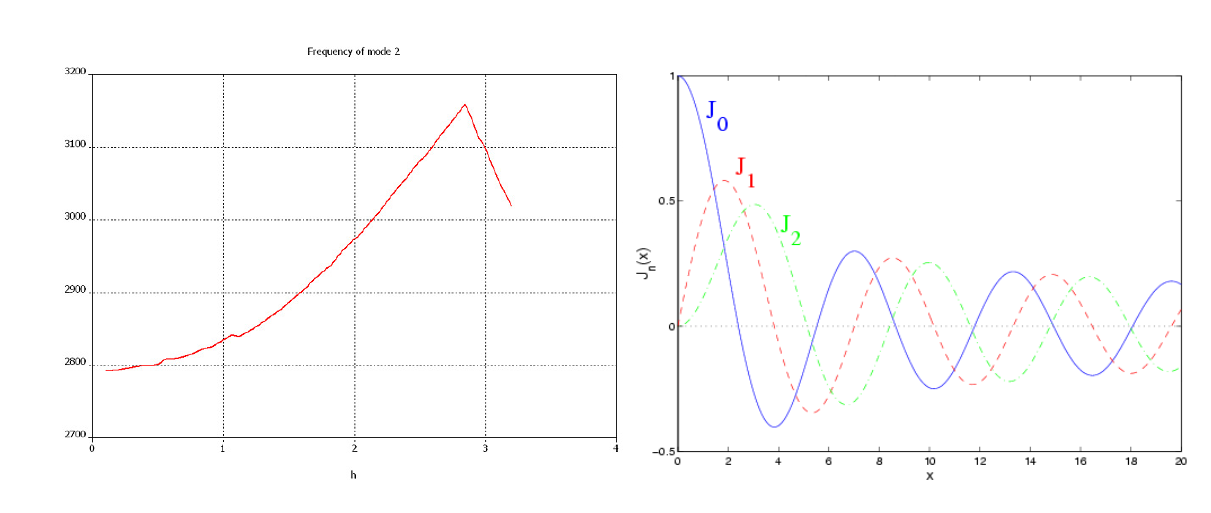}
     \label{plot9}
\caption{The frequency changes depending on the radius of the iris hole.}
\end{figure*}

\begin{figure*}
  \includegraphics[width=14cm]{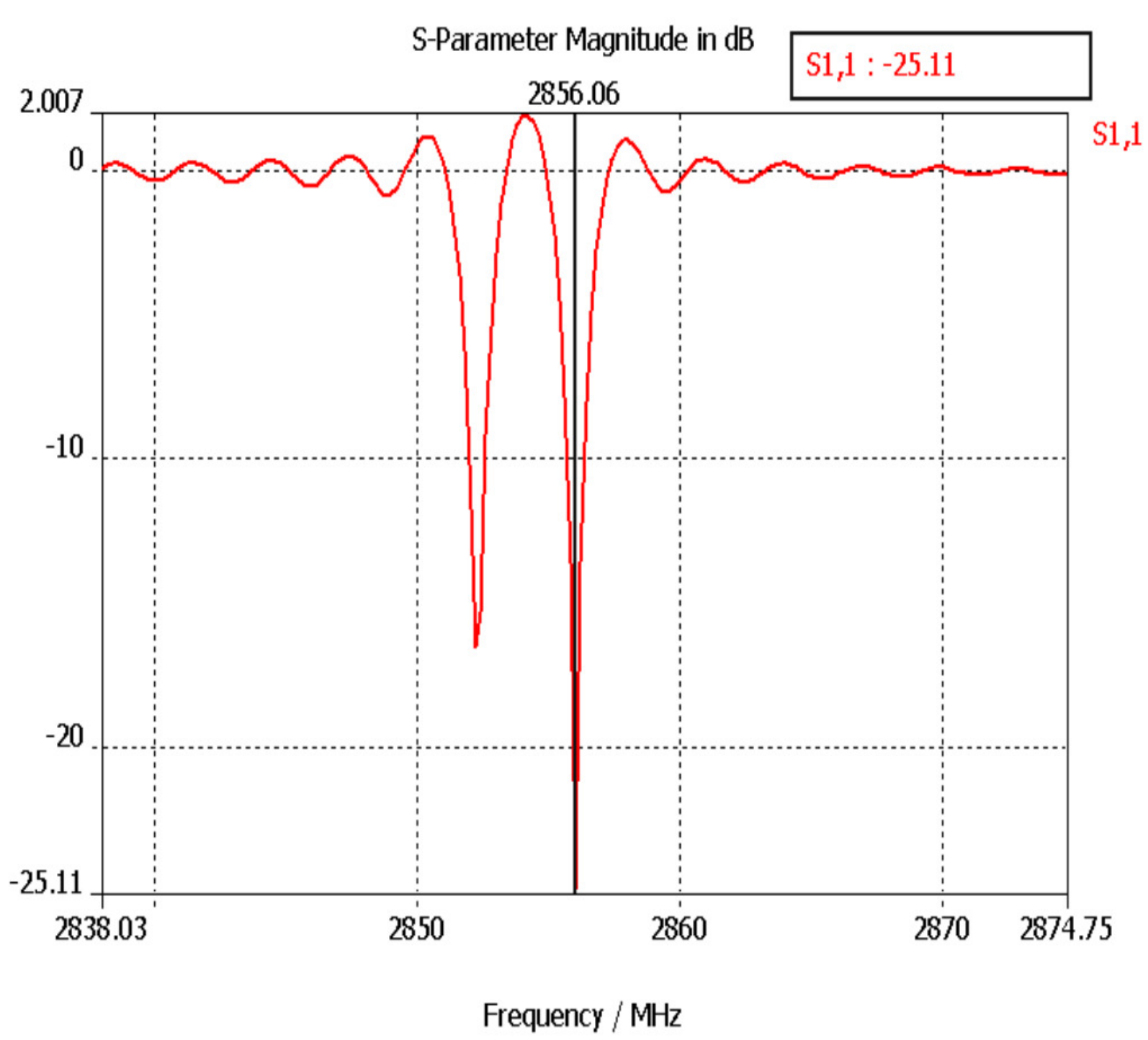}
     \label{plot10}
\caption{The graph illustrates S dB, with the $\pi$ mode occurring at 2856.06 MHz and the zero mode at 2852.2 MHz. Due to the structure's high Q value, certain frequencies show slower convergence of electromagnetic energy due to geometric properties. However, energy converges rapidly at resonant frequencies such as zero and $\pi$, as indicated by the points of decline on the graph. Consequently, despite the presence of ripples, the results remain valid.}
\end{figure*}


\begin{figure*}
    {
   \subfigure[zero mode]{
     \includegraphics[width=9 cm]{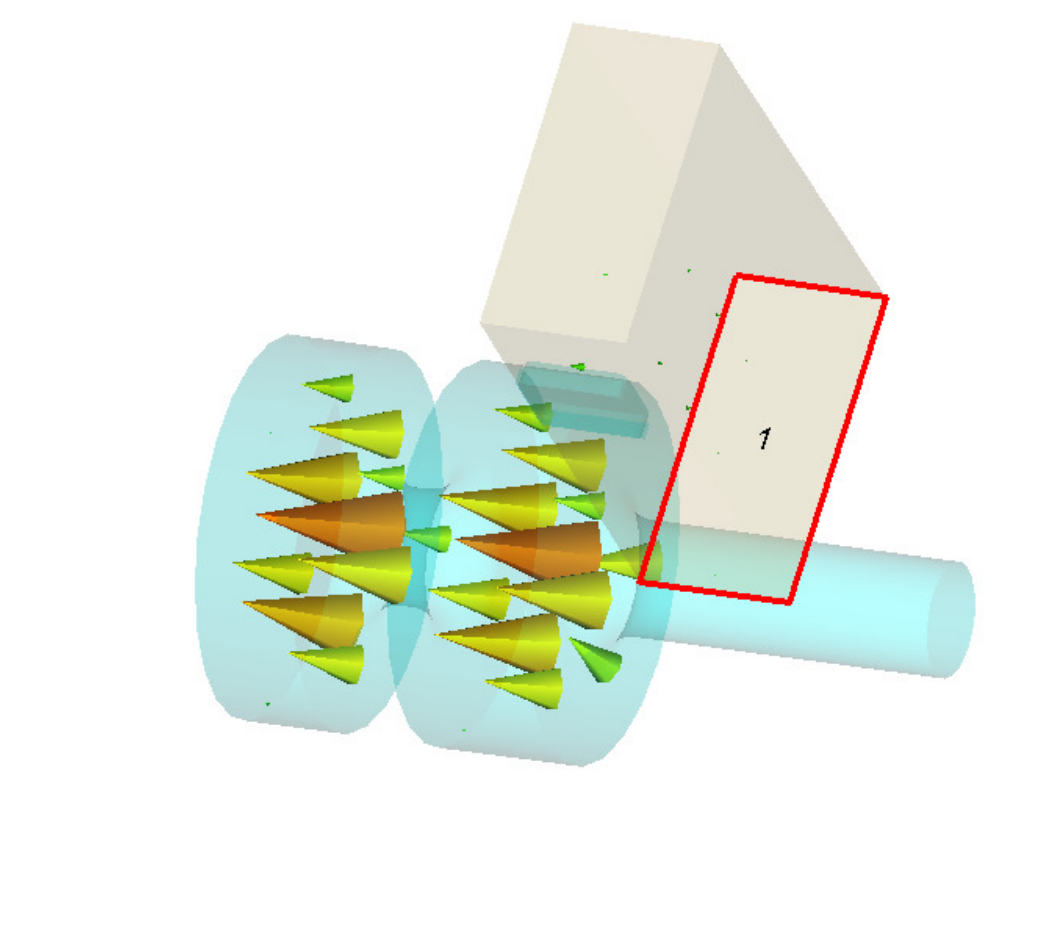}
     \label{plot11}
    }\hspace{-5 mm}
   \subfigure[$\pi$ mode]{
   \includegraphics[width=9 cm]{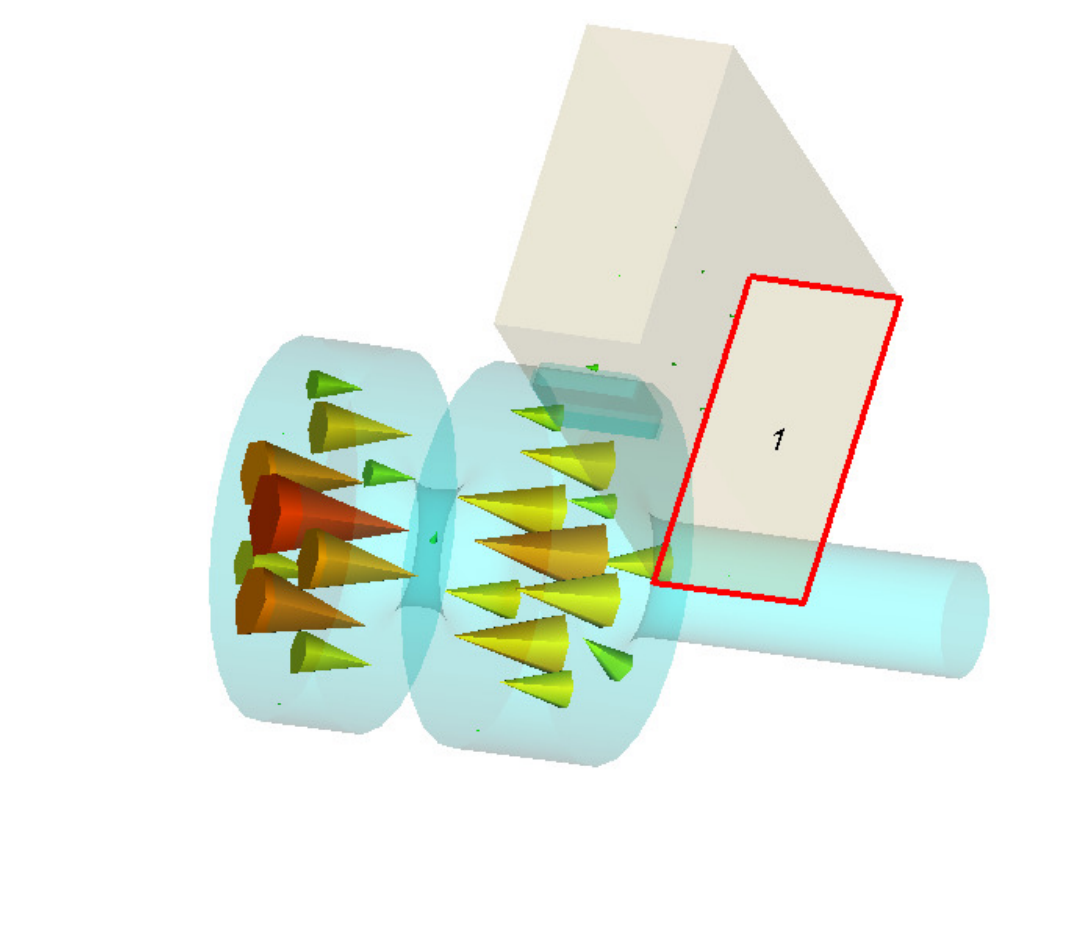}
     \label{plot12}
   }
}
\caption{The location marked with the number "1" indicates a port where the wave enters. Noticeably, there is minimal field in the waveguide, indicating that most of the energy is absorbed into the cavity. This serves as another example of weak coupling. (b) $\pi$ mode.}
\end{figure*}

\subsubsection{Equivalent circuit for two cavities}
In addition to the aperture between the waveguide and the full-length cavity, there is another aperture located on the iris positioned between the two cavities (see Fig.~\ref{plot2b}). It is imperative to account for the dipole effect at the center of the iris. As the wave propagates between the two cells, dipole sources are generated there, acting as wave generators. The principle and process are akin to the previous one. The wave mode is $TM_{010}$ (Figs.~\ref{plot3a}, \ref{plot3b}), with the longitudinal electric component ($\hat{x}$) and the tangential magnetic component ($\hat{\theta}$) taken into consideration.\\

For $TM_{01}$ mode, the space-normalized field vector that must be considered is as follows:
\begin{eqnarray}
f^{\pm} = (\mp f_{r},- f_{\theta}, f_{x}) \Rightarrow (0, -f_{\theta}, f_{x})
\end{eqnarray}
The next process is similar to the calculation of the previous one.
\begin{eqnarray}
Z_{11}(\;at \;C_{1}) = \frac{1-\chi \alpha_{x}}{i\mu \omega
(-f_{x}^{2}\alpha_{x} - f_{\theta}^{2}\alpha_{\theta} +\chi
f_{\theta}^{2}\alpha_{x} \alpha_{\theta}) }\Rightarrow \frac{1}{-i \mu \omega f_{\theta}^{2}\alpha_{\theta}  +
\frac{1}{ \frac{i}{ \mu \omega f_{x}^{2}\alpha_{x}} -
\frac{i[\chi]_{c33} }{\mu \omega f_{x}^{2} } }}.\label{impedance_Z11_3}
\end{eqnarray}
Here, $\chi$ represents the reaction tensor of the full cavity connected with the rectangular waveguide.\\

It represents the combination of impedances connected in parallel and in series. Simple substitutions are employed to derive this expression.
\begin{eqnarray}
B_{\theta _{1 or 2}} &=& -{\omega \mu f_{\theta _{1 or 2}}^{2} \alpha_{\theta _{1 or 2}}},\label{B_theta}\\
X_{x_{1 or 2}} &=& \frac{1}{\omega \mu f_{x_{1 or 2}}^{2} \alpha_{x_{1 or 2}}},\label{X_x}\\
\frac{-i [\chi_{c1 or 2}]_{33}}{\omega \mu f_{x_{1 or 2}}^{2}} &=& \frac{n_{1 or 2}^{2}}{i\omega C_{1 or 2}+\frac{1}{i\omega L_{1 or 2}}}.\label{receiving_energy}
\end{eqnarray}
Although these substitutions may seem arbitrary, they are based on reasonable rules. For example, admittance is inversely proportional to the radius of the hole, while resistance behaves oppositely, among other principles. Furthermore, in the context of half cells receiving energy [Eq.~(\ref{receiving_energy})] from the full cell and subsequently sending back reaction energy, it's essential to note that a signal from one cell also affects its neighboring cell. This scenario illustrates the typical condition of coupling. This coupling phenomenon is crucial for understanding how coupled oscillations of waves occur in connected waveguides. Such intricate interactions cannot be fully explained by simple electromagnetic wave theory applied to a single cavity.\\

Our next step involves drawing equivalent circuits for the impedances provided in Eqs.~(\ref{impedance_Z11}), (\ref{impedance_Z11_3}). In addition to the variables in those equations, we need to account for the resistances of the cavities, as well as the self and mutual inductances resulting from changes in magnetic flux within the photoinjector. These new terms must be considered when drawing the circuits (Fig.~4).\\

Moreover, we must address the effect of apertures and irises that retain electromagnetic (EM) energy. The electric and magnetic energy present in the enlarged areas influences the net resonant frequency, behaving akin to a capacitor. Therefore, it is advisable to incorporate this effect into the capacitor $C'_{1 or 2}$. The capacitors of each cavity are combined considering several factors. In the case of the full cell, the net capacitor, including tuning, is depicted as shown in Fig.~4.
\begin{eqnarray}
\frac{1}{C'_{2}}&=& \frac{1}{C_{2}}+\frac{1}{C_{apt}}+\frac{1}{C_{iris}}+\frac{1}{C_{drift}}+\frac{1}{C_{tun2}}
\end{eqnarray}
While, for the half cell, it is
\begin{eqnarray}
\frac{1}{C'_{1}}&=& \frac{1}{C_{1}}+\frac{1}{C_{iris}} + \frac{1}{C_{tun1}}.
\end{eqnarray}
For the waveguide,
\begin{eqnarray}
\frac{1}{C'_{3}}&=& \frac{1}{C_{3}}+\frac{1}{C_{aptr}}.
\end{eqnarray}

%
%
%

\subsubsection{The analysis of Equivalent Circuit}
Fig.~4 depicts the circuit representing the impedance $Z_{11}$. The left circuit in the figure corresponds to the rectangular waveguide, where energy enters the full cavity (middle circuit) through the aperture. The current in each circuit represents the energy of the waves in each cell and waveguide. Additionally, this current or energy transitions to neighboring circuits through transformers. Thus, the initial step involves solving the voltages and currents in each circuit. Subsequently, we can formulate a matrix representing the voltages and currents. The remaining task is to solve this homogeneous matrix. Ultimately, resonant frequencies are determined for nontrivial solutions, elucidating the phenomenon of coupled oscillations from the obtained results.\\

The subsequent step involves solving the equivalent circuit. This circuit is resolved utilizing Kirchhoff's loop rule. Given that the waveguide and cavities are interconnected and exert influence on each electromagnetic flux, the currents in each inductor are
\begin{eqnarray}
\left( \begin{matrix} I_{L_{1}} \cr I_{L_{2}} \cr  \end{matrix}\right)
=\frac{1}{i\omega (L_{1} L_{2} - M_{12}^{2})} \left( \begin{matrix} L_{2} & -M_{12} \cr -M_{12} & L_{1} \cr \end{matrix} \right)  \left( \begin{matrix} V_{1} \cr V_{2} \cr \end{matrix} \right),\quad
\left( \begin{matrix} I_{L_{2}} \cr I_{L_{3}} \cr \end{matrix}\right)
=\frac{1}{i\omega (L_{2} L_{3} - M_{23}^{2})} \left(
\begin{matrix} L_{3} & -M_{23} \cr -M_{23} & L_{2} \cr \end{matrix} \right)  \left(
\begin{matrix} V_{2} \cr V_{3} \cr \end{matrix} \right)
\end{eqnarray}
Here we use the definition
\begin{eqnarray}
&&\frac{L_{2}}{L_{1} L_{2} - M_{12}^{2}} = \frac{1}{L'_{1}}, \quad
\frac{L_{1}}{L_{1} L_{2} - M_{12}^{2}} = \frac{1}{L'_{2,half}},\quad
\frac{L_{3}}{L_{2} L_{3} - M_{23}^{2}} = \frac{1}{L'_{2,guide}}, \\
&&\frac{L_{2}}{L_{2} L_{3} - M_{23}^{2}} = \frac{1}{L'_{3}},\quad
\frac{M_{12}}{L_{1} L_{2} - M_{12}^{2}} = \frac{1}{M'_{12}}, \quad
\frac{M_{23}}{L_{2} L_{3} - M_{23}^{2}} = \frac{1}{M'_{23}}.\label{M12_M23_2}
\end{eqnarray}
Since the amounts of current in each node are conserved
\begin{eqnarray}
&&I_{1}= \frac{1}{i\omega L'_{1}}V_{1} - \frac{1}{i\omega M'_{12}}V_{2} + i\omega C'_{1} V_{1}
+G_{1}V_{1}+iB_{\theta_{1}}V_{1} + \frac{V_{1} - V_{2} /n_{1}}{i X_{1x}}\Rightarrow  n_{1}I'_{2} \equiv n_{1}\bigg(\frac{V_{2}-n_{1}V_{1}}{i X_{2x}}\bigg)\\
&&I_{2}= \frac{1}{i\omega L'_{2,guide}}V_{2} + \frac{1}{i\omega L'_{2,half}}V_{2} - \frac{1}{i\omega M'_{23}}V_{3} - \frac{1}{i\omega M'_{12}}V_{1} + i\omega C'_{2} V_{2} + G_{2}V_{2}+ iB_{\theta_{2}}V_{2} + \frac{V_{2} - n_{1} V_{1}}{i X_{2x}}\nonumber \\
&&\quad \Rightarrow \frac{n_{2}(V_{3} - n_{2} V_{2})}{iX_{3z}}+\frac{V_{1}-V_{2}/n_{1}}{i \, n_{1} X_{1x}}\\
&& \frac{V_{0}}{Z_{0}} = \frac{1}{i\omega L'_{3}}V_{3} - \frac{1}{i\omega M'_{23}}V_{2} + i\omega C'_{3} V_{3}+\frac{V_{3} - n_{2} V_{2}}{iX_{3z}}\\
\end{eqnarray}
The above equations are simplified as
\begin{eqnarray}
\left( \begin{matrix} Y_{11} & Y_{12}  & 0 \cr Y_{21} & Y_{22} & Y_{23} \cr  0 & Y_{32} & Y_{33}\cr \end{matrix} \right)\left(
\begin {matrix} V_{1} \cr V_{2} \cr V_{3}\cr \end{matrix} \right)= \left( \begin{matrix} 0 \cr 0 \cr I_{0}=V(0)/Z(0) \cr \end{matrix} \right),
\end{eqnarray}
where we used
\begin{eqnarray}
Y_{11}&=& i\omega C'_{1}+ \frac{1}{i\omega L'_{1}} +G_{1} + \frac{{n_{1}}^{2}}{iX_{2x}}+iB_{\theta_{1}} + \frac{1}{iX_{1x}},\quad
Y_{12}= -\frac{1}{i \omega M'_{12}} - \frac{n_{1}}{iX_{2x}}-\frac{1}{i n_{1}X_{1x}}\\
Y_{21}&=& -\frac{1}{i \omega M'_{12}} - \frac{n_{1}}{iX_{2x}}-\frac{1}{i n_{1}X_{1x}},\quad
Y_{22}= i\omega C'_{2}+ \frac{1}{i\omega L'_{2,guide}}+\frac{1}{i\omega L'_{2,half}} + G_{2} +
iB_{\theta_{2}} + \frac{1}{iX_{2x}} + \frac{n_{2}^{2}}{i X_{3z}}+\frac{1}{i n_{1}^{2} X_{1x}}\label{M12}\\
Y_{23}&=& -\frac{1}{i \omega M_{23}} - \frac{n_{2}}{i X_{3z}},\quad
Y_{32}= -\frac{1}{i \omega M_{23}} - \frac{n_{2}}{i X_{3z}},\quad Y_{33} = i\omega C'_{3}+ \frac{1}{i\omega L_{3}} + \frac{1}{iX_{3z}},\quad \frac{1}{L'_{2}} =  \frac{1}{ L'_{2,guide} } + \frac{1}{ L'_{2,half}}.\label{M23}
\end{eqnarray}
`$n^{2}_{1}/ iX_{ix}$' can be regarded as transformers through which the wave energy passes. Consequently, they may be viewed as capacitors that store electric energy.
\begin{eqnarray}
\frac{n_{i}^{2}}{i X_{jx}} \equiv i \omega C_{i}.
\end{eqnarray}
The term $1/iX_{1x}$ resembles the dimension of susceptance; however, since they are sufficiently small, we employ approximations for each term.\\

If we assume $\omega \sim \omega_{i}$
\begin{eqnarray}
i\omega C'_{i} + \frac{1}{i\omega L'_{i}} = i\omega C'_{i}(1  -
\frac{\omega_{i}^{2}}{\omega^{2}}) \sim 2 i C'_{i} (\omega -
\omega_{i}),
\end{eqnarray}
we obtain a simplified matrix as follows:
\begin{eqnarray}
\left( \begin{matrix} Y'_{11} & Y'_{12}  & 0 \cr Y'_{21} & Y'_{22} & Y'_{23} \cr  0 & Y'_{32} & Y'_{33}\cr \end{matrix}
\right)\left( \begin{matrix} V_{1} \cr V_{2} \cr V_{3}\cr \end{matrix} \right)= \left( \begin{matrix} 0 \cr 0 \cr I_{0} \cr \end{matrix} \right),
\end{eqnarray}
where the definitions are
\begin{eqnarray}
Y'_{11}&=&2i\,(C'_{1}+C_{1,n})(\omega - \omega_{1}) + G_{1} + iB'_{\theta_{1}},\\
Y'_{12}&\simeq &  -\frac{1}{i \omega M'_{12}} = Y'_{21} \equiv -\kappa_{12}\label{kappa12},\\
Y'_{22}&=&2i\,(C'_{2}+C_{2,n})(\omega - \omega_{2}) + G_{2} + iB'_{\theta_{2}},\\
Y'_{23}&\simeq& -\frac{1}{i \omega M'_{23}} = Y'_{32} \equiv -\kappa_{23}\label{kappa23},\\
Y'_{33}&=&2i \, C'_{3}(\omega - \omega_{3}) + iB'_{3z}.
\end{eqnarray}
If we observe above terms carefully, we find the diagonal terms have
similar components.  And, we use the relation
\begin{eqnarray}
-\omega + \omega_{j} - \frac{G_{j} -
iB_{\theta_{j}}}{2i(C'_{j}+C_{j,n})} &=& -\omega + \omega_{j}
+i\tau_{j}.\label{time_constant}
\end{eqnarray}
We get the matrix as follows:
\begin{eqnarray}
\left( \begin{matrix} -\omega + \omega_{1} + i\tau_{1} & -\kappa_{12}  & 0 \cr -\kappa_{12} & -\omega + \omega_{2} +
i\tau_{2} & -\kappa_{23} \cr 0 & -\kappa_{23} & -\omega + \omega_{3} + i\tau_{3}\cr \end{matrix} \right)\left( \begin{matrix}
V_{1} \cr V_{2} \cr V_{3}\cr \end{matrix} \right)= \left( \begin{matrix} 0 \cr 0 \cr I_{0}/(-2iC'_{3}) \cr \end{matrix} \right).
\end{eqnarray}
The above inhomogeneous matrix presents a typical challenge that requires diagonalization. As each voltage is interconnected with its neighboring term, various coupled modes such as "$\pi$" or "zero" emerge. To address this matrix, the initial step involves determining eigenvalues and their corresponding eigenvectors. Subsequently, the inhomogeneous term $I_{0}$, representing the driven wave at 2.856 GHz, is introduced into the system. Due to the ohmic loss of the copper photoinjector, the resonance of the driven solution does not precisely occur at 2.856 GHz. By analyzing the voltage amplitude differentiation, the exact resonant frequency can be identified. However, this aspect is omitted in this thesis as it holds less significance compared to other variables in computer simulation and can be adjusted during the tuning of a real photoinjector. The primary objective is to obtain eigenvalues, various frequencies, and their corresponding mode behaviors. Among these modes, the focus of this thesis is to determine the dimension of the "pi" mode frequency at 2.856 GHz and its associated well-balanced electric field.\\

The general solution of this matrix is rather long and complex. However, if the matrix satisfies three conditions, it becomes easily calculable. The first requirement is that the resonance frequency of the $\pi$ mode is 2.856 GHz. Second, the field amplitudes of the full cell and cavity 2 (or the 0.6 cell) are nearly identical but with opposite phases ($\pi$ radian). The third requirement is that the amplitude of $V_{3}$ is zero, ensuring the corresponding S value is also zero. This condition is crucial for absorbing all energy carried by the wave into the photoinjector, rendering the electric field in the rectangular waveguide nonexistent. With these conditions established, we proceed to solve the matrix.\\

Considering the third condition, the voltage amplitude of $V_{3}$ in the rectangular waveguide should approach zero. Achieving this necessitates a weak coupling between the full cell and the rectangular waveguide. While some specialized cases are discussed in mechanics texts \cite{thornton2004classical, 2002clme.book.....G}, many real coupled oscillators do not satisfy such specific conditions. When the geometrical resonant frequencies of the rectangular waveguide and the full cell differ, a coupling coefficient $\kappa_{23}$ much smaller than the difference in resonant frequencies leads to an eigenvector ratio close to zero. This type of coupling, known as 'weak coupling,' implies that energy is predominantly stored in one side rather than being shared evenly. Thus, the aperture's shape and size that meet this condition must be determined to ensure all wave energy enters the full cell. Conversely, energy in the injector should be evenly distributed between both cells (full cell and 0.6 cell), requiring strong coupling between them. As the geometrical resonant frequencies of both cells also differ, the coupling constant $\kappa_{12}$ must exceed the difference in frequencies.
\begin{eqnarray}
|\Omega_{2}- \Omega_{3}| \gg \kappa_{23} \qquad (\Omega_{j}= \omega_{j} + i \tau_{j}). \cr
\end{eqnarray}
And the second requirement implies a strong coupling between the two cavities:
\begin{eqnarray}
&&|\Omega_{1}-\Omega_{2}| \leq \kappa_{12}.
\end{eqnarray}
Then, the eigenvalue and eigenvector for the zero mode are
\begin{eqnarray}
\omega_{0} &=& \frac{-\kappa^{2}_{23} + \Omega_{3}(\Omega_{1}+\Omega_{2}) - \sqrt{\Theta_{1}^{2} - 4
\Theta_{2}}} {2\Omega_{3} }\simeq \frac{\Omega_{1}+\Omega_{2}-\sqrt{4\kappa_{12}^{2}+(\Omega_{1}-\Omega_{2})^{2}}}{2}.\label{freq. of zero mode}
\end{eqnarray}
Then, the amplitudes are
\begin{eqnarray}
(V_{1},V_{2}) &=& (\kappa^{2}_{23} + \Omega_{3}(\Omega_{1}-\Omega_{2}) - \sqrt{\Theta_{1}^{2} - 4
\Theta_{2}},\; - 2\kappa_{12}\Omega_{3})\cr &\sim &(-(\Omega_{1}-\Omega_{2}) +
\sqrt{4\kappa_{12}^{2}+(\Omega_{1}-\Omega_{2})^{2}},
 \; 2\kappa_{12})\cr &\sim& (1 : 1).\label{amplitude of zero mode}
\end{eqnarray}
For $\pi$ mode,
\begin{eqnarray}
\omega_{\pi} &=& \frac{-\kappa^{2}_{23} + \Omega_{3}(\Omega_{1}+\Omega_{2}) + \sqrt{\Theta_{1}^{2} - 4
\Theta_{2}}} {2\Omega_{3} } \simeq \frac{\Omega_{1}+\Omega_{2} + \sqrt{4\kappa_{12}^{2}+(\Omega_{1}-\Omega_{2})^{2}}}{2},\label{freq. of Pi mode}
\end{eqnarray}
and
\begin{eqnarray}
(V_{1},V_{2}) &=& (\kappa^{2}_{23} +
\Omega_{3}(\Omega_{1}-\Omega_{2}) + \sqrt{\Theta_{1}^{2} - 4
\Theta_{2}},\; -2\kappa_{12}\Omega_{3}) \cr &\sim &
(-(\Omega_{1}-\Omega_{2})-\sqrt{4\kappa_{12}^{2}+(\Omega_{1}-\Omega_{2})^{2}},
\; 2\kappa_{12}) \cr &\sim & (-1 : 1). \label{amplitude of pi mode}
\end{eqnarray}
Here we used the definition:
\begin{eqnarray}
\Theta_{1} = \kappa_{23}^{2}-\Omega_{3}(\Omega_{1}+\Omega_{2}),\quad \Theta_{2} =
\Omega_{3}(-\kappa_{23}^{2}\Omega_{1}-\kappa_{12}^{2}\Omega_{3}+ \Omega_{1}\Omega_{2}\Omega_{3}).
\end{eqnarray}
And the third mode for the rectangular waveguide is
\begin{eqnarray}
\omega &=& \omega_{3},\quad (V_{1},V_{2},V_{3}) \sim (0,0,1)
\end{eqnarray}
The eigenvector sets for higher and lower frequency,
\begin{eqnarray}
|u^{+}\rangle &\sim&  \left( \begin{matrix}\beta - \sqrt{4\kappa_{12}^{2}+\beta^{2}} \cr 2\kappa_{12}\cr\end{matrix}
\right)\qquad (\beta \equiv \Omega_{1}- \Omega_{1}),\\
|u^{-}\rangle &\sim&  \left( \begin{matrix}\beta +
\sqrt{4\kappa_{12}^{2}+\beta^{2}} \cr 2\kappa_{12}\cr \end{matrix} \right) \sim
\left( \begin{matrix}-2\kappa_{12}\cr  \beta -
\sqrt{4\kappa_{12}^{2}+\beta^{2}} \cr \end{matrix} \right).
\end{eqnarray}
The amplitude ratio of the above results is approximately 1. If $\beta$, which is related to the difference between both geometrical frequencies, is small, we achieve an equally balanced '$\pi$' and 'zero' mode amplitude.

\section{Simulation Results}
\subsection{Eigenmode Properties and Dimensions of Each Cell}
Figs.~5, 6, 7 depict the simulation results. Additionally, theoretically calculated values are provided for comparison. In this analysis, the computer program 'Microwave Studio' was utilized, known for producing highly accurate results. However, discrepancies exist between the two sets of values. This discrepancy is attributed to round-off errors in computation, as the theoretical equations are known to be exact (derived from Electromagnetism). The method for calculating capacitance and reactance is given by:
\begin{eqnarray}
C_{i} = \frac{4W}{V^{2}} = \frac{\pi\epsilon R_{i}^{2} |J_{0}'(p_{01})|^{2}}{\ell_{i}},\quad L_{i} = \frac{4W}{I^{2}} = \frac{\mu \ell_{i}  \int_{0}^{R_{i}}  rJ_{1}(kr) dr } {2\pi
R_{i}^{2}J_{1}(kR_{i}^{2}) }.\label{C and L}
\end{eqnarray}
Fig.~8 illustrates that the mode resonant frequency is inversely proportional to the radius. However, several factors influence the frequency. $\Omega$ in Eq.~(\ref{freq. of Pi mode}) comprises the geometrical frequency and time constant. According to Eqs.~(\ref{C and L}), $C_{j}$ is proportional to $R_{j}^{2} / \ell_{j}$, and $L_{j}$ depends on $R_{j}$. Conversely, Eq.~(\ref{time_constant}) shows that the time constant $\tau_{j}$ depends on the radius and is affected by resistance, aperture or iris size. Additionally, the relative field amplitudes of each cell change according to Eq.~(\ref{amplitude of pi mode}). The coupling constant $\kappa$ decides the relative amplitude of the electric field, and even this constant depends on the reactance or radius of each cavity [refer to Eqs.~(\ref{kappa12}), (\ref{kappa23}), (\ref{M12}), (\ref{M23}), (\ref{M12_M23_2})].\\

Fig.~9 illustrates the variation in the resonant frequency as the radius of the iris increases. Unlike the cavity radius, the response is quite distinctive.As the size of the iris increases, the magnitude of the H-field, which is smaller than that of the E-field near the axis, becomes larger. Both quantities are functions of Bessel functions. When the iris size is 2.8 cm, the change in the magnitude of H and E field energies is the same. Furthermore, impedance or susceptance terms such as $X_{i}$ or $B_{i}$ in $\Omega_{i}$ depend on the size of the iris [Eqs.~(\ref{B_theta}), (\ref{X_x}), (\ref{time_constant})]. With these composite elements, the pattern of resonant frequency changes as depicted in Fig.~9. Moreover, as the equations derived indicate, the aperture's effect plays a role in the resonant frequency and field amplitude but in an opposite manner.\\

\subsection{Simulation results of transient mode and dimensions of the photoinjector}

\subsubsection{Dimension of 1.6 cell photoinjector}
Figs.~11 depict the results of a transient mode simulation, while Figs.~5, 6, 7 are derived from an eigenmode simulation. The latter simulations were conducted without setting a port where waves are generated. As the wave modes are solely determined by the boundary conditions of the photoinjector geometry, they may deviate significantly from real wave mode behaviors. Hence, it is imperative to simulate the photoinjector with a port. In Figs.~11, waves enter through a port marked with the number '1', and nearly all of the field energy enters the cavity through the aperture. Fig.~11(a) displays the electric field shape of the '0' mode, while Fig.~11(b) illustrates the electric field for the 'pi' mode, signifying a phase angle difference of $\pi$ radians. Fig.~10 indicates that two modes occur at frequencies of 2852.2 MHz and 2856.06 MHz. Furthermore, the S dB of the '$\pi$' mode is -25.11, implying that approximately 9 out of 10 units of field energy are absorbed.
\begin{table}
\centering
\begin{tabular}{|c|c|c|c|c|c|c|c|} \hline
\multicolumn{2}{|c|} {0.6 Cell} & \multicolumn{2}{c|} {full cell} & \multicolumn{2}{c|} {iris} & \multicolumn{2}{c|} {Drift} \\
\hline \hline radius & length & radius & length & radius & length &  radius & length \\ \hline 4.157 & 2.27964
& 4.189 & 3.2512 & 0.9525 & 0.29972 & 0.93 & 7.265 \\ \hline
\end{tabular}
\caption{unit: cm}
\end{table}

\section{Summary}
To summarize, in terms of energy, the phase velocity of the wave in a single cell is faster than $c$, so it cannot store enough energy to accelerate the beam bunch effectively. Consequently, a multi-celled cavity is employed to overcome these limitations. By introducing holed irises in a cylindrical waveguide, the phase velocity becomes slower than $c$, enabling sufficient energy storage to accelerate electron beams rapidly in a compact photoinjector. Nonetheless, as the structure becomes more complex and the volume of the waveguide changes, both the field and resonant frequency also undergo alterations. These phenomena are elucidated using the concept of perturbation theory. However, the theory alone cannot explain the harmonic oscillation of the field, which is a crucial aspect. To address this, it is necessary to convert the wave signal into equivalent currents.\\

In our work, we have referred to the analytic approaches of \citep{1995PhDT........39L}. However, this model is tailored for a 17GHz photoinjector and overlooks the process of energy transfer through the holed iris. Since omitting energy transfer through the holed iris is inaccurate, certain modifications are required. Simulation results confirm that if there is no energy transfer through the holed iris, the field exists only in the cavity. Additionally, the rectangular waveguide is attached differently. Also, their model forms Electric and Magnetic dipoles, but the Electric dipole poses a risk of voltage breakdown. Moreover, the calculations are intricate, making it challenging to avoid nonlinear terms. Consequently, many approximations, which lack clear physical justification, are inevitable. However, attaching the rectangular waveguide as depicted in Fig.~2 can address these issues. As detailed in section 3, the new model mitigates the risk of voltage breakdown by excluding the electric dipole source at the aperture, while also simplifying calculations and ensuring precise calculated impedances. Following the construction of the circuit from the impedance, the analysis results elucidate the shift in resonant frequency and field amplitude. Some approximations for minor variables and assumptions regarding couplings among structures have been made. Based on the trends of these variables, simulation results have been obtained. The next step in the future involves fabricating the real cavity and comparing experimental results with computational ones.\\

Our analytical calculations elucidate the functions of the zero mode frequency, $\pi$ mode frequency, and the corresponding amplitudes. These are dependent on parameters such as $\omega$, $\tau$, and the coupling constant $\kappa$. The circuit variables are determined by the impedance Z ($=1/Y$), capacitance C, and mutual conductance M. It is crucial to determine the physical variables, such as radius and length, corresponding to these circuit parameters. However, the most critical aspect lies in comprehending the physical significance of the (weak) coupling among the injector cells as physical oscillators.

\begin{acknowledgements}
K. Park acknowledges the support from National Research Foundation of Korea:NRF-2021R1I1A1A01057517.
\end{acknowledgements}

\bibliography{bibfile_2024_0426}

\end{document}